\documentclass[aps,prb,twocolumn,groupedaddress]{revtex4-2}

\usepackage{graphicx}
\usepackage{amsmath}
\usepackage{xcolor}

\usepackage[breaklinks]{hyperref}
\hypersetup{
    pdfstartview={FitH},
    colorlinks=true,
    linkcolor=blue,
    citecolor=blue,
    urlcolor=blue
}

\begin{document}

\title{An investigation of the sensitivity of the Fermi surface to the treatment of exchange and correlation}

\author{E.~I.~Harris-Lee}
\affiliation{H.\ H.\ Wills Physics Laboratory,
University of Bristol, Tyndall Avenue, Bristol, BS8 1TL, United Kingdom}

\author{A.~D.~N.~James}
\affiliation{H.\ H.\ Wills Physics Laboratory,
University of Bristol, Tyndall Avenue, Bristol, BS8 1TL, United Kingdom}

\author{S.~B.~Dugdale}
\affiliation{H.\ H.\ Wills Physics Laboratory,
University of Bristol, Tyndall Avenue, Bristol, BS8 1TL, United Kingdom}

\date{\today}

\begin{abstract}

The Group V and VI transition metals share a common Fermi surface feature of hole ellipsoids at the $N$ point in the Brillouin zone.
In clear contrast to the other Fermi surface sheets, which are purely of $d$ character, these arise from a band that has a significant proportion of $p$ character.
By performing local density approximation (LDA), generalized gradient approximation (GGA), strongly constrained and appropriately normed (SCAN) meta-GGA, and $GW$ approximation calculations, we find that the $p$ character part of this band (and therefore the Fermi surface) is particularly sensitive to the exchange-correlation approximation.
LDA and GGA calculations inadequately describe this feature, predicting $N$ hole ellipsoid sizes that are consistently too large in comparison to various experimental measurements, whereas quasiparticle self-consistent $GW$ calculations predict a size that is slightly too small (and non-self-consistent $GW$ calculations that use an LDA starting point predict a size that is much too small).
Overall, for the metals tested here, SCAN provides the most accurate Fermi surface predictions, mostly correcting the discrepancies between measurements and calculations that were observed when LDA calculations were used.
However, none of the tested exchange-correlation approximations succeeds in simultaneously bringing all of the measurable properties of these metals into good experimental agreement, particularly where magnetism is concerned. The SCAN calculations predict antiferromagnetic moments for Cr that are 3 times larger than the experimental value (1.90 $\mu_B$ compared to 0.62 $\mu_B$).

\end{abstract}


\maketitle

\section{Introduction}

Density-functional theory (DFT) has been remarkably successful, both qualitatively and quantitatively, in predicting the Fermi surfaces of a wide range of metallic systems, these predictions being tested with a range of experimental techniques \cite{dugdale_life_2016}. 
Yet the element Cr presents a small but notable exception to this success.
Bulk Fermi surface measurements using the de Haas-van Alphen (dHvA) effect \cite{graebner_haas-van_1968,laurent_band_1981}, two-dimensional angular correlation of positron annihilation radiation (2D-ACAR) \cite{dugdale_fermiology_1998}, and Compton scattered photons \cite{tanaka_study_2000,dugdale_high-resolution_2000} all consistently produce a Fermi surface that is notably different from the existing theoretical predictions, including DFT calculations with the local density approximation (LDA) \cite{laurent_band_1981,kubler_spin-density_1980}.
A small, similar disagreement has been noted for V \cite{laurent_energy_1978,singh_electron_1985}, but the match between experiment and theory is much better for other elemental transition metals, including Mo \cite{dugdale_fermiology_1998,sparlin_empirical_1966}.

It is obviously desirable to understand the origin of any discrepancies properly, and to correct them if it possible to do so.
Experimental origins cannot be ruled out, but are less likely to be responsible when different experimental techniques, each with different sets of strengths and flaws, highlight similar discrepancies with the same Fermi surface feature.
The existing evidence therefore suggests that the problem is more likely to be with the theoretical calculations.
More specifically, the exchange-correlation approximations that have been employed so far (mostly LDA) are likely to be at fault.
The Fermi surface is known to be a sensitive test of the quality of an exchange-correlation approximation \cite{nickerson_prediction_1976}, and, moreover, one of the earliest bandstructure calculations on Cr, which used a `Hartree-Fock-Slater' treatment of the exchange-correlation potential, noted the complete absence of the $N$ hole ellipsoids for Cr \cite{loucks_fermi_1965}.
Therefore there is prior theoretical evidence that this particular Fermi surface feature, being the main source of inconsistency between experiment and theory, is especially sensitive to the treatment of exchange and correlation.
If it is the exchange-correlation approximation at fault, these discrepancies could be a useful opportunity to explore how and why the most widely used exchange-correlation approximations can start to fail, even outside materials where strong electron correlations are expected to exist such as electronically complex oxides.
Another possibility is that it is not a good enough approximation to use nonmagnetic calculations to represent the Fermi surfaces of the magnetic phases in which the measurements have actually been made.
This is an especially plausible factor in Cr, which displays particularly complex magnetic behaviour that can be connected to its Fermi surface geometry \cite{fawcett_spin-density-wave_1988}.
At this time, there are still notable problems with theoretical predictions of the magnetic behaviour of Cr.
For example, neither the paramagnetic nor the incommensurate antiferromagnetic spin-density wave (SDW) state has actually been found to be the magnetic groundstate when the LDA is used \cite{pindor_disordered_1983,hafner_magnetic_2002}.

The intention of this paper is to gather and link the results of various existing experiments to find consistent discrepancies between experiment and theory. 
Next, we attempt to understand and explain these with the aid of new theoretical calculations; in particular, we have performed DFT calculations with the strongly constrained and appropriately normed (SCAN) meta-generalized gradient approximation (MGGA) \cite{sun_strongly_2015}, and various $GW$-approximation calculations (single-step perturbation \cite{godby_self-energy_1988} and `quasiparticle self-consistent' \cite{van_schilfgaarde_adequacy_2006}).
We have also explored the role of the magnetic state in determining the Fermi surface of Cr.

\section{Basic Fermi surface geometry}

\begin{figure}
	\includegraphics[width=0.25\textwidth]{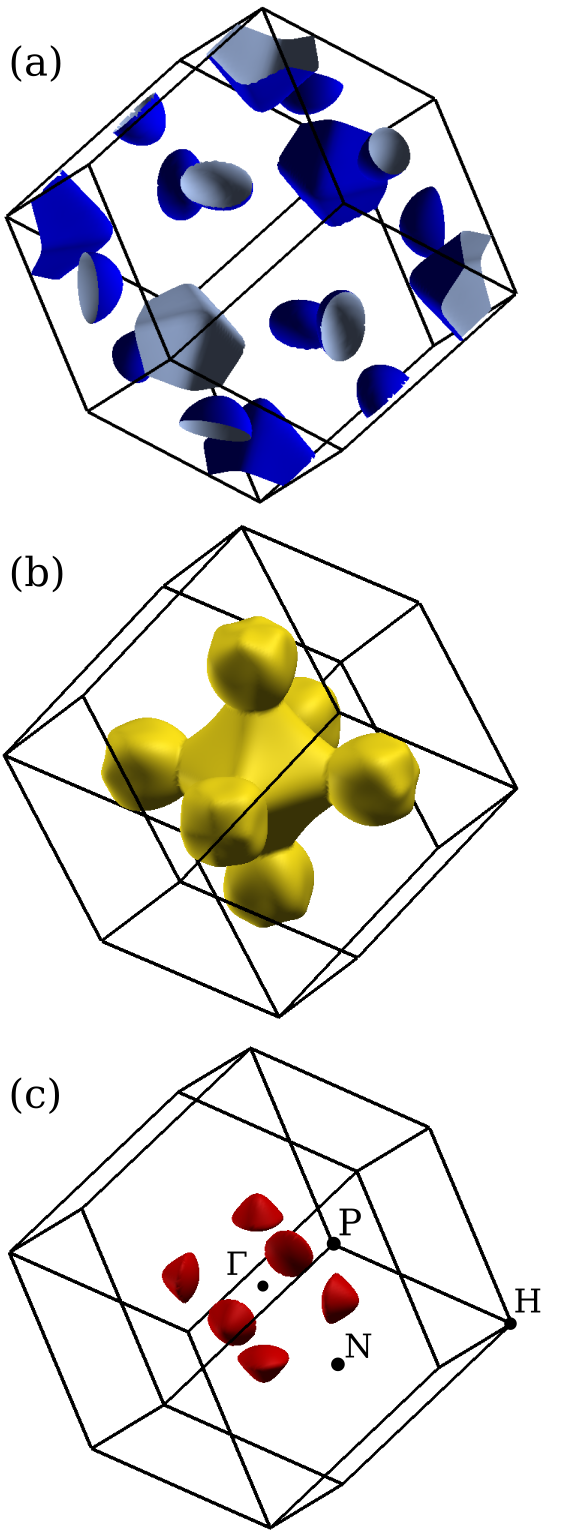}
	\caption{\label{fig:CrMoW_FS} The Fermi surface that is attributed to the Group VI metals. Three Fermi surface sheets are displayed. (a) hole pockets surrounding the $N$ and $H$ points (ellispoidal at $N$, octahedral at $H$). (b) the `jack' composed of an electron octahedron centered on $\Gamma$ and  electron `balls' that are centered along the ${\Gamma}H$ path. (c) the electron `lenses', which are positioned along the ${\Gamma}H$ high symmetry lines where the octahedra and balls overlap. 
	}
\end{figure}
The Fermi surface geometry of the Cr-group (Group VI) metals was first established theoretically by Lomer \cite{lomer_electronic_1962}. 
Despite Lomer's use of a very simple model (extrapolating the Fermi surface by changing the band filling for the bandstructure of Fe), DFT-LDA calculations do produce Fermi surfaces with essentially the same features and geometry; only the sizes of these features differ \cite{laurent_band_1981,kubler_spin-density_1980}.
This Fermi surface consists of one hole sheet and two electron sheets (four distinct levels of electron occupation in total) in the body centered cubic (BCC) Brillouin zone, as is shown in Fig.~\ref{fig:CrMoW_FS} (this is a DFT-LDA prediction and a corresponding DFT-LDA bandstructure can be found in the Results, in Sec.~\ref{sec:bandstr_nonmag} Fig.~\ref{fig:Cr_bandstr_lspec}).
There are five distinct features: the $N$ centered hole ellipsoids and $H$ centered hole octahedra comprise the lowest, hole-like occupation level; the first electron-like occupation level consists of the `jack', which is a combination of the $\Gamma$ centered electron octahedra and the electron balls; the overlap between these $\Gamma$ octahedra and balls results in the final, highest occupation feature, which are the electron lenses.

The V-group (Group V) transition metals also form a BCC crystal structure, and share the $N$ hole ellipsoid Fermi surface feature with the Cr-group (Group VI) metals.
The full V-group Fermi surface geometry is shown in Ref.~\cite{parker_experimental_1974}.

All of these terms (octahedra, ellipsoid) are, of course, approximate descriptions of the shapes.
For example, the term `ellipsoid' appears to be a slightly poorer description of the shapes of the $N$ hole pockets in the V-group, where these may be elongated along the ${\Gamma}N$ lines \cite{parker_experimental_1974}.

\section{Existing experimental measurements}

There is an important distinction between Cr and the adjacent elemental metals that is relevant to their Fermi surfaces: at very low temperature Cr is antiferromagnetic, exhibiting a longitudinally polarized, incommensurate spin-density wave of a length that is expected to match the Fermi surface nesting vector between the $H$ and $\Gamma$ octahedra \cite{fawcett_spin-density-wave_1988,heil_accurate_2014}. 
At T$_S$~=~121~K this SDW becomes transversely polarized, and at higher temperatures still (T$_N$~=~310~K), Cr becomes paramagnetic. 
The neighbouring elemental metals, including V, Nb, Mo and W, do not exhibit magnetic order (remaining paramagnetic) \cite{fawcett_spin-density-wave_1988}.
This information about the magnetic behaviour of these metals has been obtained from extensive neutron scattering studies \cite{shull_neutron_1953,arrott_neutron-diffraction_1967,fincher_magnetic_1979,fawcett_spin-density-wave_1988}.

There are several well-established techniques for measuring the Fermi surface of a metal, each with distinct advantages and disadvantages.
A comprehensive picture of the Fermi surface can often only be obtained through analysis of the results of more than one of these different experimental techniques.

Many of the earliest experimental measurements of Fermi surfaces were performed by exploiting the dHvA effect \cite{shoenberg_magnetic_1984}.
This method for obtaining the bulk Fermi surface geometry can be extremely accurate, but it requires certain conditions such as high magnetic field, long electron mean free path, and low temperature. 
As a result, for Cr this type of measurement is restricted to the Fermi surface of the longitudinal antiferromagnetic SDW state. 
The results of dHvA measurements on Cr \cite{graebner_haas-van_1968} have previously been compared to an LDA calculation by Laurent \textit{et al.} \cite{laurent_band_1981}, who found that the largest source of disagreement between the measurements and their calculation was the $N$ hole ellipsoids; these were found to be smaller than predicted.
(At this point it is worth noting that agreement is not improved when the more recent electronic structure codes are used.
When combined with current computational power these might be thought to provide greater precision.
In addition to there being disagreement over the size of the $N$ hole ellipsoids, our new LDA and GGA calculations also demonstrate an electron ball size disagreement that is markedly larger than was previously noted. See Sec.~\ref{sec:dHvA} for more detail.)
In V the measured $N$ hole ellipsoids are also smaller than predicted by LDA calculations \cite{laurent_energy_1978}.
In contrast, dHvA measurements of the $N$ hole ellipsoids of paramagnetic Mo \cite{sparlin_empirical_1966,hoekstra_determination_1973} demonstrate reasonable agreement with theoretical predictions, including some of the earliest ones \cite{loucks_fermi_1965,koelling_fermi_1974}.
These Mo measurements provide an important reference for the level of experiment-theory agreement that should be achievable.
It should be noted that an explanation based on experimental difficulties cannot be entirely precluded, especially for Cr because interpretation of the Cr measurements involved an unusual extra step: to compare these results to theoretical calculations in the BCC Brillouin zone, it was necessary to create a model BCC Fermi surface that best fit the experimental data (which has tetragonal symmetry because of the SDW).
But this model Fermi surface has been tested by multiple authors, who have been unable to find an alternative model that better matches the experimental data overall \cite{reifenberger_electron_1980}.
No such transformation from tetragonal to cubic symmetry was necessary for V, since it has a simpler magnetic state at very low temperature (it is paramagnetic), yet the sizes of the $N$ hole ellipsoids are overpredicted for V too.
We also note that, as their name suggests, the $N$ hole ellipsoids are conveniently situated at the $N$ high symmetry points, do not have a particularly complicated geometry, and have relatively low effective masses, so produce strong measurement signals.
Even without looking at the results of measurements with alternative experimental techniques, it therefore seems unlikely that the $N$ hole ellipsoid experiment-theory discrepancies are mainly due to experimental artifacts or other experimental difficulties.

Experimental measurements of the Fermi surface of Cr, Mo, and V have also been performed by the 2D-ACAR technique \cite{kaiser_positron_1987,dugdale_fermiology_1998,hughes_evolution_2004,singh_electron_1985}. 
Strictly, it is the momentum distribution of the electrons as seen by the positron that is measured by this technique, from which the occupation density, and thereby the Fermi surface, can be obtained \cite{dugdale_probing_2014}.
This method only measures 2D projections of the Fermi surface (although the full 3D Fermi surface can be reconstructed tomographically from a series of projections). 
However, it is not restricted to low temperatures, which means that bulk measurements of the Fermi surface of both the paramagnetic and antiferromagnetic SDW states of Cr can be performed.
The $N$ high symmetry points are projected onto themselves along the [110] projection direction, so measurement along this direction yields a clear picture of the $N$ hole ellipsoids.
The paramagnetic state 2D-ACAR measurements (the sample was heated to T~=~353~K, well above T$_N$) show $N$ hole ellipsoid sizes that are smaller than indicated by the DFT-LDA prediction, just as measurements of the antiferromagnetic SDW do \cite{fretwell_reconstruction_1995,dugdale_fermiology_1998,hughes_evolution_2004}. 
It is more difficult to draw conclusions about other individual Fermi surface features, because these overlap with each other in any projection, but the overall agreement can be compared.
There is a small but notable difference between dHvA results and LDA calculations for Cr, but the difference between 2D-ACAR measurements and LDA calculations is conspicuous (for both the $N$ hole ellipsoids and the rest of the Brilluoin zone). 
Contrastingly, the 2D-ACAR measurements on Mo and V are consistent with dHvA results - for Mo the measured Fermi surface is in good agreement with LDA theoretical predictions \cite{kaiser_positron_1987,dugdale_fermiology_1998}, and for V there is a small but noteworthy difference in the size of the $N$ hole ellipsoids \cite{singh_electron_1985}.
Several attempts were made to determine the cause of the very large differences between LDA calculations and 2D-ACAR measurements on Cr.
For example, one line of investigation followed from the observation that Cr experiment-theory agreement could be improved by convolution of the theoretical data with a Gaussian function (one much broader than the well-known experimental resolution would require) to represent the momentum distribution smearing that can sometimes be induced by strong many-body correlations \cite{matsumoto_two-dimensional_1986,migdal_momentum_1957,luttinger_fermi_1960,daniel_momentum_1960,barbiellini_treatment_2001,olevano_momentum_2012,hiraoka_direct_2020}.
A more complete argument follows the reasoning that the excessive disagreement could be mainly caused by strong positron-electron interactions in Cr, which could have influenced the measurement \cite{matsumoto_many-body_1987,dugdale_fermiology_1998,laverock_experimental_2010}.
In an attempt to address this possibility, Compton scattering measurements \cite{tanaka_study_2000,dugdale_high-resolution_2000} were performed; this method is similar to 2D-ACAR in the sense that it is the electron momentum distribution that is measured, but here using the inelastic scattering of high energy X-rays, with which there is no doubt that the groundstate electronic structure remains unperturbed \cite{cooper_compton_1985,dugdale_probing_2014}.
The disadvantages of Compton scattering are that the experimental momentum resolution is worse than it is for 2D-ACAR measurements, and that only the 1D (twice projected) electron momentum distribution can be directly measured, from which a 2D or 3D distribution must be reconstructed.
Ultimately these Compton scattering measurements were found to support the conclusion that the $N$ hole ellipsoids are smaller than DFT-LDA calculations predict  \cite{tanaka_study_2000,dugdale_high-resolution_2000}.
This means that although the positron does appear to have an unusually large, undesirable influence on the measured electron momentum density in Cr, which hampers a straightforward visualization of the Fermi surface, there still remains a genuine underlying experiment-theory discrepancy that is congruent with that revealed by the dHvA measurements.

Angle resolved photoemission spectroscopy (ARPES) measurements on Cr films do exist \cite{rotenberg_electron_2005,rotenberg_surface_2008}.
The surface and bulk electronic structure of Cr are known to be quite different \cite{zabel_magnetism_1999}, so the likeness of bulk and surface Fermi surfaces is not guaranteed.
Ignoring any concerns about the bulk representability of these ARPES measurements, it seems that the $\Gamma$ and $H$ octahedra both compare well with LDA calculations \cite{rotenberg_electron_2005,rotenberg_surface_2008}. 
ARPES Fermi surface measurements on Mo films are also available \cite{jeong_experimental_1989,kroger_angle-resolved_2000}.
Overall, as far as we can tell, this ARPES data for Cr and Mo is consistent with the dHvA and 2D-ACAR data that is the main focus of this paper.

Kohn-anomaly measurements can be made to obtain nesting vector lengths between various Fermi surface sheets, and therefore obtain some limited information about the Fermi surface geometry \cite{shaw_investigation_1971}.
We choose not to focus on comparisons to experimental Kohn-anomaly measurements of Fermi surface nesting vectors here because these offer neither the high precision nor full picture of the Fermi surface that is afforded by other techniques.
One very useful piece of information that can be obtained from these measurements is that Kohn anomalies do not change significantly from the spin-density-wave to paramagnetic state, which indicates that the Fermi surface does not change significantly either, or at least that the parts responsible for the anomalies do not change.

\section{Theoretical Methods}

\subsection{Motivation: $GW$ calculations}

In $GW$ approximation calculations the exchange-correlation density-functional (that is used in DFT) is replaced by a non-local `self-energy'.
This self-energy is calculated from first principles, constructed using a Green function ($G$) and a dynamically screened Coulomb interaction ($W$) \cite{hedin_new_1965}.
The $GW$ approximation is essentially the Hartree-Fock approximation plus the most important part of the beyond-Hartree-Fock correlations for describing weak and moderately correlated systems realistically, which is added in the form of screening \cite{aryasetiawan_gw_1998,onida_electronic_2002,reining_gw_2018}.
There are several different versions of the $GW$ approximation \cite{hybertsen_electron_1986,von_barth_self-consistent_1996,bruneval_effect_2006,van_schilfgaarde_adequacy_2006}.
The simplest is probably the one where a single-step (`one-shot') perturbation is applied to a reasonable starting point, which is normally a converged DFT calculation \cite{hybertsen_electron_1986,godby_self-energy_1988}.
We refer to this as the $G^{LDA}W^{LDA}$ method (when the starting point is a DFT-LDA calculation). 
Originally the use of this non-self-consistent $GW$ method was purely motivated by computational expense, but it is now often argued that the one-shot $GW$ approximation can be better justified than various versions of $GW$ that do incorporate a level of self-consistency \cite{onida_electronic_2002,marom_benchmark_2012,schone_self-consistent_1998}.
In some cases, however, introducing a degree of self-consistency can be vitally important \cite{gatti_understanding_2007,vanschilfgaarde_quasiparticle_2006}.
The fact that errors arise in some self-consistent schemes is related to the fact that the $GW$ approximation is an interacting one (it is not possible to find a set of noninteracting states, whereas it is in DFT or the Hartree-Fock approximation).
The `quasiparticle self-consistent' $GW$ method, referred to here as QSGW, avoids the typical $GW$ self-consistency problems by making the best possible noninteracting approximation to the interacting $GW$ self-energy.
This method removes the unsystematic dependency on the starting point that is present in one-shot calculations, but without introducing the large errors that are found in some types of $GW$ self-consistency \cite{van_schilfgaarde_adequacy_2006}.
In this paper we present both $G^{LDA}W^{LDA}$ and QSGW calculations.

In the Results Section we will note that the band that is responsible for the $N$ hole ellipsoid Fermi surface is predominantly $p$-like near to the $N$ point (Sec.~\ref{sec:bandstr_nonmag}, Fig.~\ref{fig:Cr_bandstr_lspec}).
This is important because the $GW$ approximation to the self-energy can sometimes be insufficient for describing very localized $d$ and $f$ states, but it is expected to be highly accurate for $s$ and $p$ states, and is most famous for realistic predictions of the bandgaps of semiconductors and insulators, many of which have dominant $s$ and $p$ character valence and conduction states \cite{aryasetiawan_gw_1998,vanschilfgaarde_quasiparticle_2006,shishkin_self-consistent_2007}.
Calculations that employ the $GW$ approximation may therefore provide a more realistic description of the Fermi surfaces (and general electronic structure) of the elemental Group V and VI metals, because the main discrepancy is a Fermi surface feature with $p$ character.

\subsection{Motivation: DFT calculations with SCAN}

On the ladder of increasingly complex exchange-correlation approximations, MGGA density-functionals sit above LDAs and GGAs \cite{perdew_jacobs_2001,sala_kinetic-energy-density_2016}, but are below approximations that incorporate exact exchange (e.g. hybrid density-functionals \cite{becke_new_1993,perdew_rationale_1996}) and those that incorporate exact exchange and partial exact correlation (e.g. $GW$ \cite{hedin_new_1965}, which is not a density-functional approximation, but fits the definition `exact exchange and partial exact correlation').
At this time, MGGAs are actively being designed, tested, and improved.

A significant, recently developed MGGA is the SCAN density-functional \cite{sun_strongly_2015}, which is named both for the fact that it is the first fully constrained MGGA functional (it satisfies all of the 17 known exact mathematical constraints that a MGGA should), and for the fact that it is constructed to be exact or near-exact for a range of `appropriate norms' (these include uniform electron densities and appropriately chosen atoms).
The SCAN functional yields more accurate predictions than the PBE (Perdew-Burke-Ernzerhof \cite{perdew_generalized_1996}) GGA for some material properties \cite{sun_strongly_2015,shahi_accurate_2018,zhang_competing_2020,furness_accurate_2018,pulkkinen_coulomb_2020}, but is less accurate for others \cite{fu_applicability_2018,fu_density_2019,isaacs_performance_2018,mejia-rodriguez_analysis_2019,tran_shortcomings_2020}.
The SCAN functional has been found to generally improve the predictions of semiconductor bandgaps compared to GGA and LDA functionals, although these are usually still underpredicted \cite{yang_more_2016,jana_assessing_2018}.
Many of these insulators have predominant $s$ and $p$ character, so it is plausible that SCAN may improve the Fermi surface prediction of Cr, since the discrepancy is associated with a Fermi surface feature with $p$ character (see Fig.~\ref{fig:Cr_bandstr_lspec} in Results Sec.~\ref{sec:bandstr_nonmag}).

\subsection{Technical details}
\label{sec:technical}

We have used experimental lattice constants (a~=~2.884~\r{A} for Cr, a~=~3.147 \r{A} for Mo, and a~=~3.024~\r{A} for V \cite{wyckoff_crystal_1963}) unless stated otherwise.
We found that including spin-orbit coupling does not alter the Mo bandstructure in a significant way, so we do not include it in the results presented here.

To perform DFT calculations (including SCAN calculations) we primarily used the full-potential Elk code \cite{noauthor_elk_nodate},  with a linearized augmented plane wave (LAPW) plus local-orbital basis.
The Elk code \cite{noauthor_elk_nodate} is capable of performing self-consistent SCAN MGGA calculations, in conjunction with Libxc \cite{lehtola_recent_2018}.
In some versions of Elk (including the most recent version at the time of writing), the regularized SCAN (RSCAN) functional \cite{bartok_regularized_2019} must be used instead of SCAN; we have checked this functional and find that the SCAN and RSCAN results for the Fermi surfaces are consistent.
We used a version of Elk that has been modified (extended) to be capable of calculating electron momentum densities \cite{ernsting_calculating_2014} and electron-positron momentum densities \cite{weber_spin-resolved_2015}.

The electron momentum density can be calculated from the momentum-space wave functions and the occupation numbers of each state. For a DFT calculation the electron momentum density can be expressed
\begin{equation}
\rho(\mathbf{p}) = \sum_{\mathbf{k},j} n_{\mathbf{k},j} \left| \int \psi_{\mathbf{k},j} (\mathbf{r}) e^{-i\mathbf{p}\cdot\mathbf{r}} d\mathbf{r} \right|^2,
\end{equation}
where $\psi_{\mathbf{k},j} (\mathbf{r})$ is the real-space Kohn-Sham wavefunction with wavevector $\mathbf{k}$ and state $j$, and $n_{\mathbf{k},j}$ is its occupation.
Our implementation uses a linear tetrahedron method to perform the necessary interpolations \cite{matsumoto_improvement_2004,ernsting_calculating_2014}.
The electron momentum density as it is seen by a positron (for comparison to 2D-ACAR experimental data) can be expressed \cite{weber_spin-resolved_2015}
\begin{equation}
\rho^{e - p}(\mathbf{p}) = \sum_{\mathbf{k},j} n_{\mathbf{k},j} \left|  \int \psi_{\mathbf{k},j} (\mathbf{r}) \psi_{+}(\mathbf{r})  \sqrt{\gamma(\mathbf{r})} e^{-i\mathbf{p}\cdot\mathbf{r}} d\mathbf{r} \right|^2,
\end{equation}
where $\psi_{+}$ is the positron wavefunction and $\gamma(\mathbf{r})$ is the positron enhancement factor, which takes account of electron-positron correlations beyond the independent particle model \cite{drummond_quantum_2011,laverock_experimental_2010,barbiellini_calculation_1996,mitroy_enhancement_2002,jarlborg_local-density_1987}.

The low number of valence electrons (relative to the number in many of the calculations that are now routinely performed) meant that it was straightforward to obtain well converged calculations for the nonmagnetic and commensurate antiferromagnetic states.
Extra caution was given to the fact that calculations of the electron momentum density (Sec.~\ref{sec:2D-ACAR}) require a dense grid of k-points: we used at least 400 k-points in the irreducible BZ to ensure satisfactory convergence of these \cite{ernsting_calculating_2014}.
For electron momentum density calculations the momentum cutoff was 18 a.u. (or 9 a.u. for calculations in which the positron influence is included, since the positron wave function has vanishingly small overlap with the most localized electrons which contribute at the highest momenta).
We use the Drummond positron enhancement factor \cite{drummond_quantum_2011} for the calculations that include a positron influence.

To perform $GW$ calculations ($G^{LDA}W^{LDA}$ and QSGW \cite{van_schilfgaarde_adequacy_2006}) we have used the Questaal code \cite{pashov_questaal_2020}, employing a full potential (Hankel function) LMTO plus local orbital basis.

The bandstructures from Elk DFT-LDA calculations were checked against those for DFT-LDA calculations using Questaal. The level of consistency was found to be very high (0.05 eV maximum difference between bands).
All of the presented DFT results are Elk calculations, unless it is explicitly stated otherwise.

\section{Results}

\subsection{Theoretical bandstructures}

\subsubsection{Nonmagnetic}
\label{sec:bandstr_nonmag}

\begin{figure*}
	\includegraphics[width=0.95\textwidth]{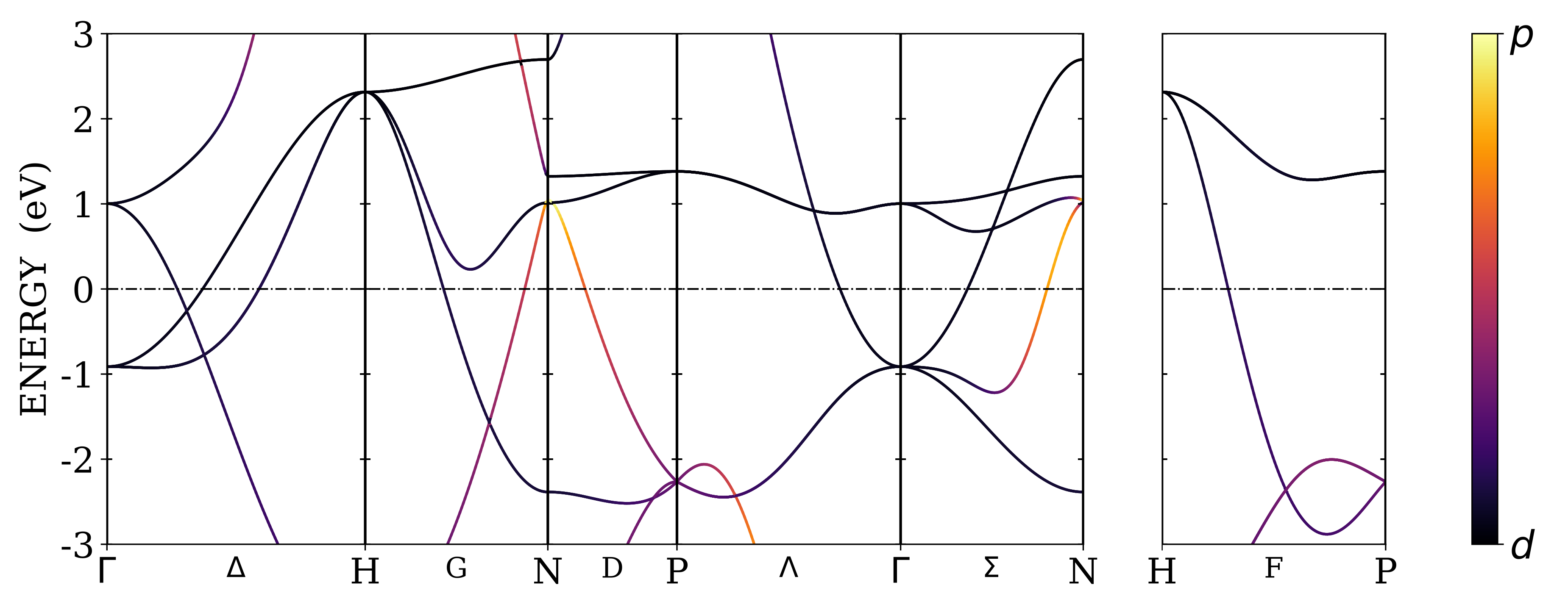}
	\caption{\label{fig:Cr_bandstr_lspec} Nonmagnetic Cr bandstructure according to DFT (LDA \cite{perdew_accurate_1992}) calculations, with a color scale indicating the dominant character of each band when it is decomposed according to angular momentum quantum number. The Fermi energy is at 0.}
\end{figure*}
\begin{figure*}
	\includegraphics[width=0.9\textwidth]{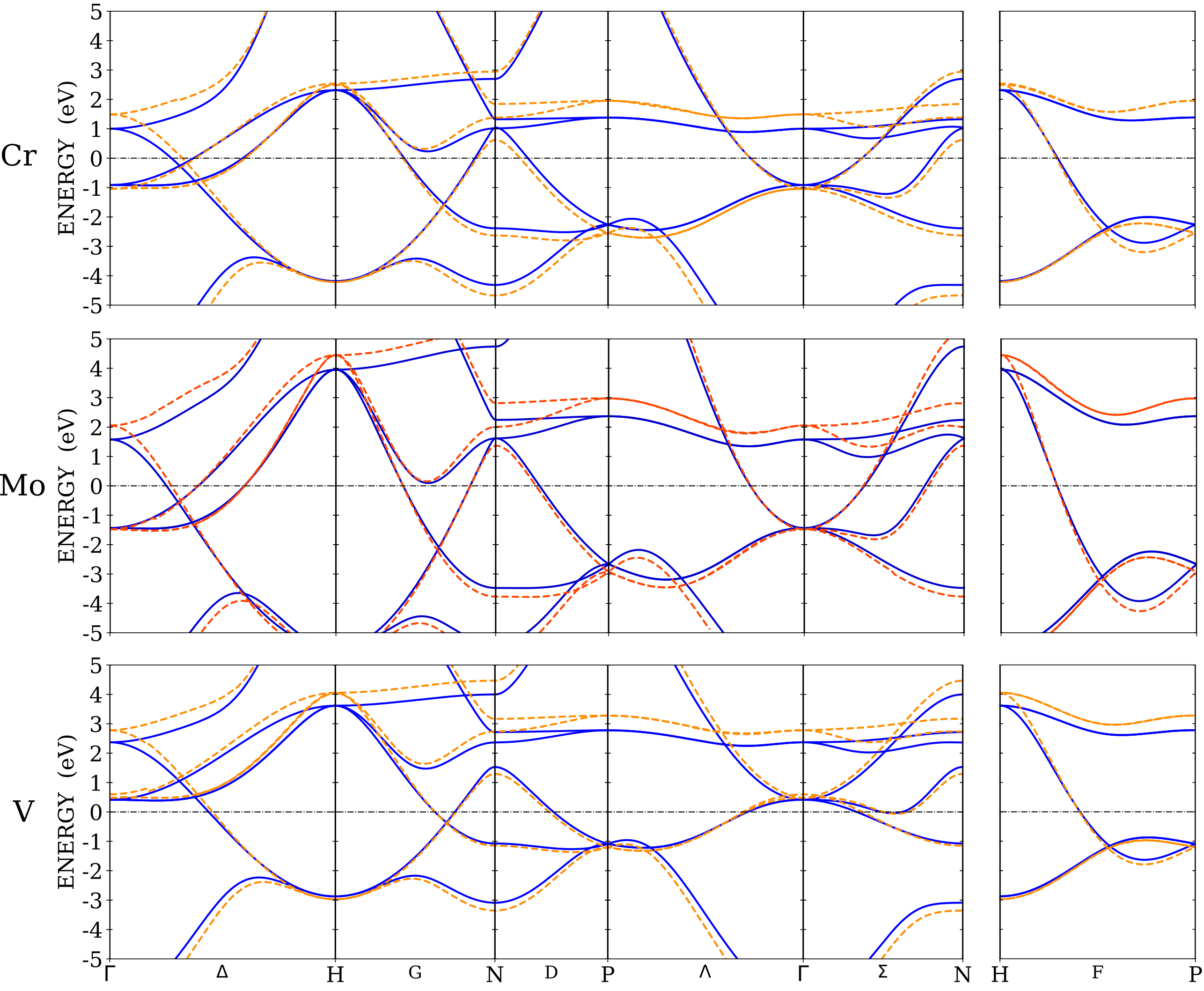}
	\caption{\label{fig:CrMoV_bandstr_SCAN}  The DFT LDA \cite{perdew_accurate_1992} (blue, solid line) and DFT SCAN \cite{sun_strongly_2015} (orange/red, dashed line) bandstructures for nonmagnetic Cr (top) and Mo (middle) and V (bottom). The Fermi energy is at 0.}
\end{figure*}

It is simplest to first consider the Fermi surfaces of these metals in the absence of magnetic order (the paramagnetic states).
For the purposes of this paper we generally make the assumption that the paramagnetic and nonmagnetic states are equivalent to a very good approximation, since proper theoretical treatment of paramagnetic behaviour is considerably more difficult \cite{pindor_disordered_1983}.

Figure~\ref{fig:Cr_bandstr_lspec} shows the bandstructure and orbital angular momentum band character of Cr for nonmagnetic LDA calculations in an energy window near to the Fermi energy.
When decomposed according to their angular momentum character, most of the bands that cross the Fermi energy have close to 100\% $d$ character.
The exception is the band that is responsible for the $N$ hole ellipsoids, which is entirely $p$ character at the $N$ point.
Details of the character of this band are listed in Table~\ref{tab:bandchar}, which also shows that these remain essentially the same for the SCAN exchange-correlation approximation, and indeed for others which are not shown here.
For V there is a higher proportion of $d$ character where this band crosses the Fermi energy, but, overall, for V, Nb, and Mo the band character decomposition is very similar to this one for Cr (in all cases the band has a significant proportion of $p$ character at the points where it crosses the Fermi energy near to the $N$ point, forming an $N$ hole ellipsoid).
\begin{table}
	\caption{\label{tab:bandchar} Table of band characters for Cr for the key band that creates the $N$ hole ellipsoids, decomposed according to angular momentum quantum number at two points in the Brillouin zone: at the $N$ point, and at the point along ${\Gamma}$N where the band crosses the Fermi energy (which is different for LDA and SCAN). Values are quoted to the nearest whole \%.}
	\begin{ruledtabular}
		\begin{tabular}{c|cccc}
			&\multicolumn{2}{c}{LDA \cite{perdew_accurate_1992}}&\multicolumn{2}{c}{SCAN \cite{sun_strongly_2015}}
			\\
			Char.&${\Gamma}$N&N&${\Gamma}$N&N
			\\
			\hline 
			$s$&4&0&3&0
			\\
			$p$&39&53&45&53
			\\
			$d$&18&0&9&0
			\\
			$f$&1&1&1&1
			\\
			unassigned&38&46&41&45
			\\
		\end{tabular}
	\end{ruledtabular}
\end{table}

In Figure~\ref{fig:CrMoV_bandstr_SCAN} the LDA bandstructure is compared to the SCAN bandstructure for nonmagnetic Cr, Mo and V. 
The bandstructures for LDA \cite{perdew_accurate_1992} and PBE GGA \cite{perdew_generalized_1996} calculations are difficult to distinguish from each other when plotted on the same axes (not shown).
The LDA and SCAN bandstructures for Cr are broadly similar, but SCAN increases the widths of the $d$ bands (which is consistent with a shift of occupied $d$ bands to lower energies that has previously been observed for SCAN calculations on other transition metals \cite{ekholm_assessing_2018}), and the $p$ character band that rises above the Fermi energy near to the $N$ point is also notably altered. 
The positions where the bands cross the Fermi energy are noticeably different for the SCAN calculations.
For Mo, the $d$ bands are also slightly wider for the SCAN calculation, but there is a smaller difference between the Fermi energy band crossings and a smaller change to the states that have $p$ character than there is for Cr.
For V the changes are consistent with Cr and Mo: the $d$ bands are slightly wider in the SCAN calculation and the $N$ hole ellipsoid sizes are reduced.

\begin{figure}
	\includegraphics[width=0.4\textwidth]{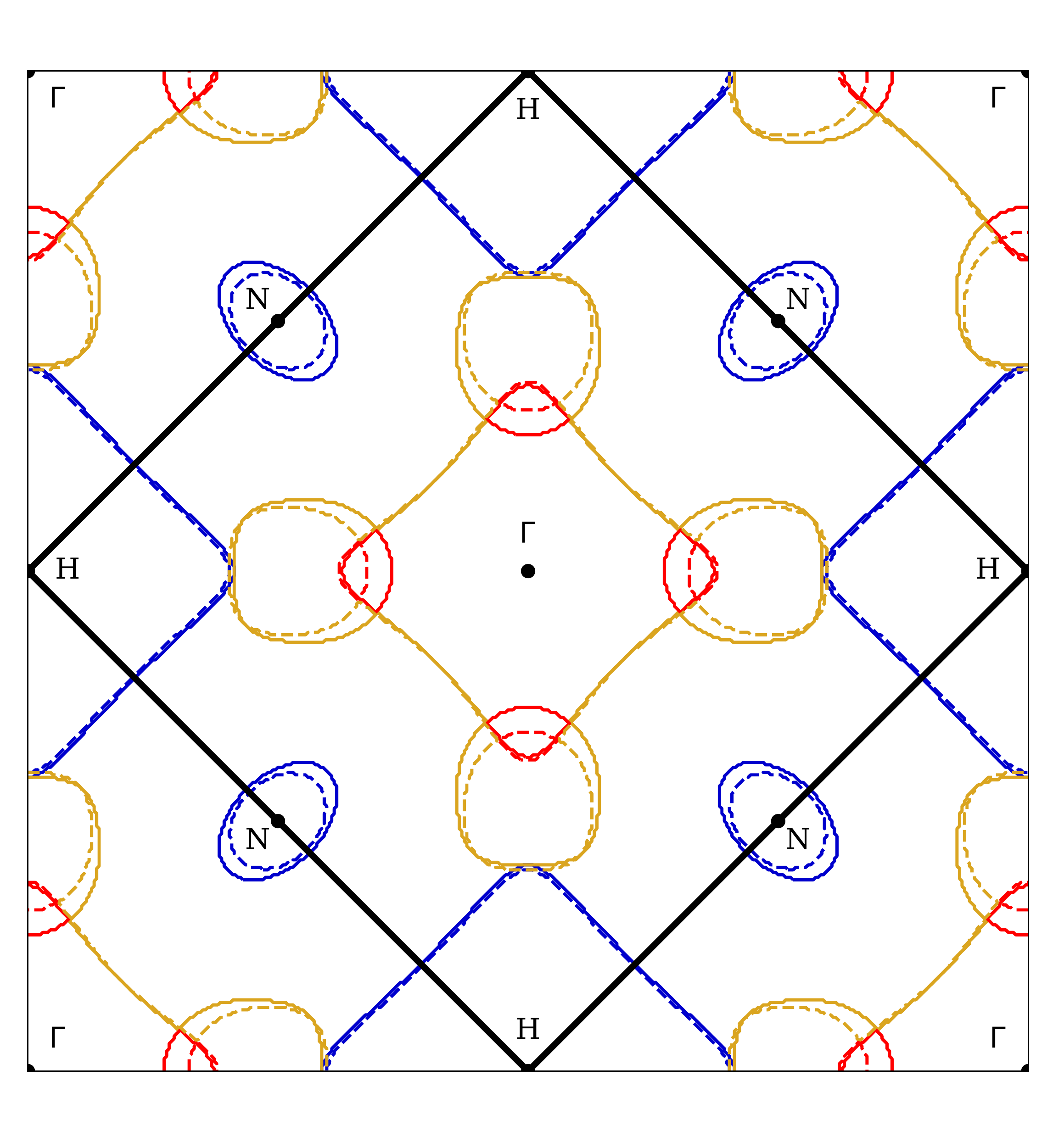}
	\caption{\label{fig:Cr_FS_plane} }
	The Fermi surface of Cr in the (001) plane that intersects the $\Gamma$ point, according to LDA (solid lines) and SCAN MGGA (dashed lines) calculations. The colors match those used in Fig.~\ref{fig:CrMoW_FS}. The boundary of the BCC Brillouin zone is indicated by the thick black line. The repeated zone scheme has been used. 
\end{figure}
\begin{figure}
	\includegraphics[width=0.45\textwidth]{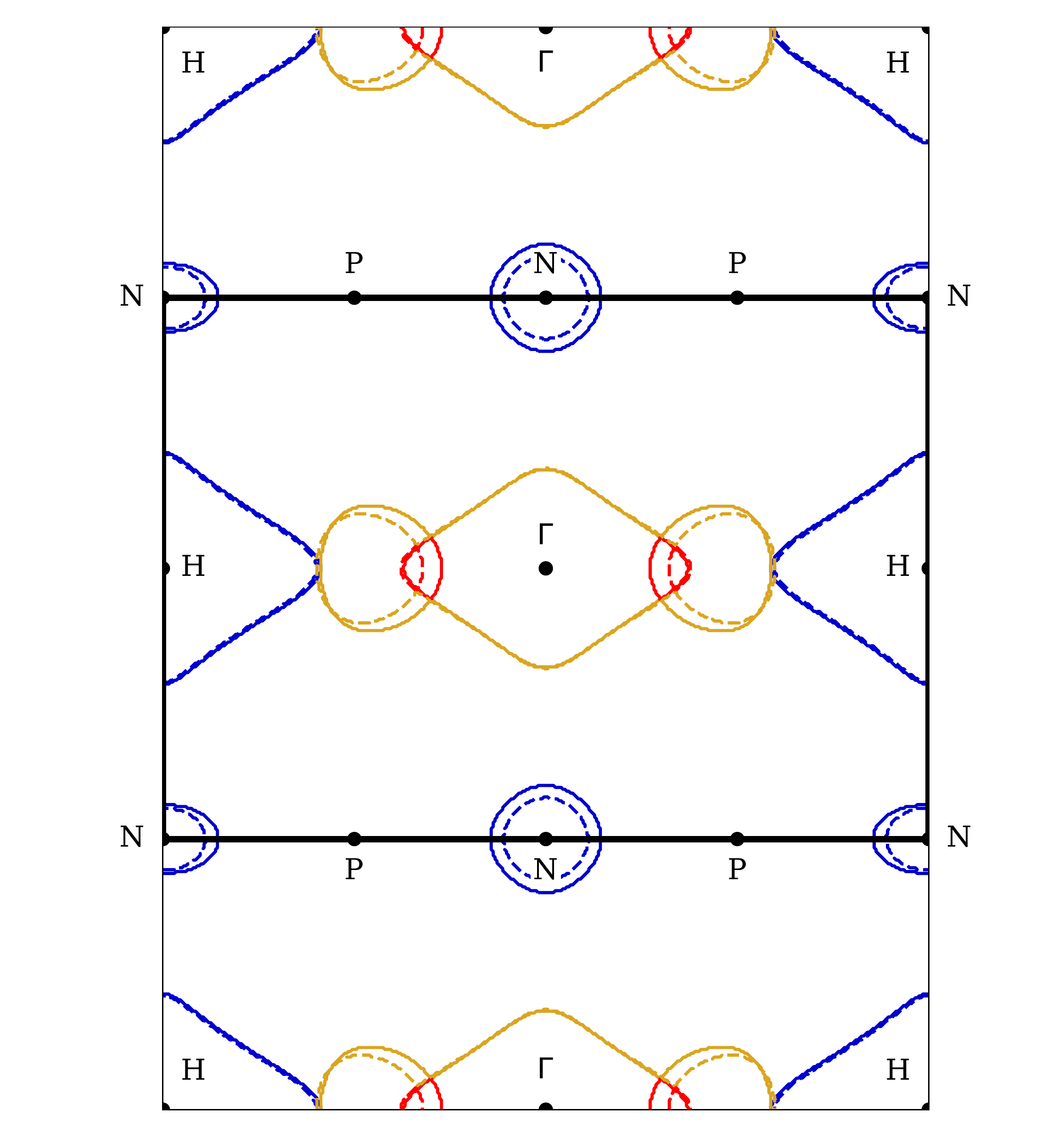}
	\caption{\label{fig:Cr_FS_plane_110} }
	The Fermi surface of Cr in the (110) plane that intersects the $\Gamma$ point, according to LDA (solid lines) and SCAN MGGA (dashed lines) calculations. The line colors correspond to those used in Fig.~\ref{fig:CrMoW_FS}. The boundary of the BCC Brillouin zone is indicated by the thick black line. The repeated zone scheme has been used. 
\end{figure}
Figure~\ref{fig:Cr_FS_plane} and Figure~\ref{fig:Cr_FS_plane_110} show the difference between the Fermi surface of Cr in (001) and (110) planes for LDA and SCAN MGGA calculations. 
The electron balls and $N$ hole ellipsoids are clearly affected by the treatment of the exchange-correlation potential, whereas the two octahedral parts of the Fermi surface are not.

\begin{figure*}
	\includegraphics[width=0.9\textwidth]{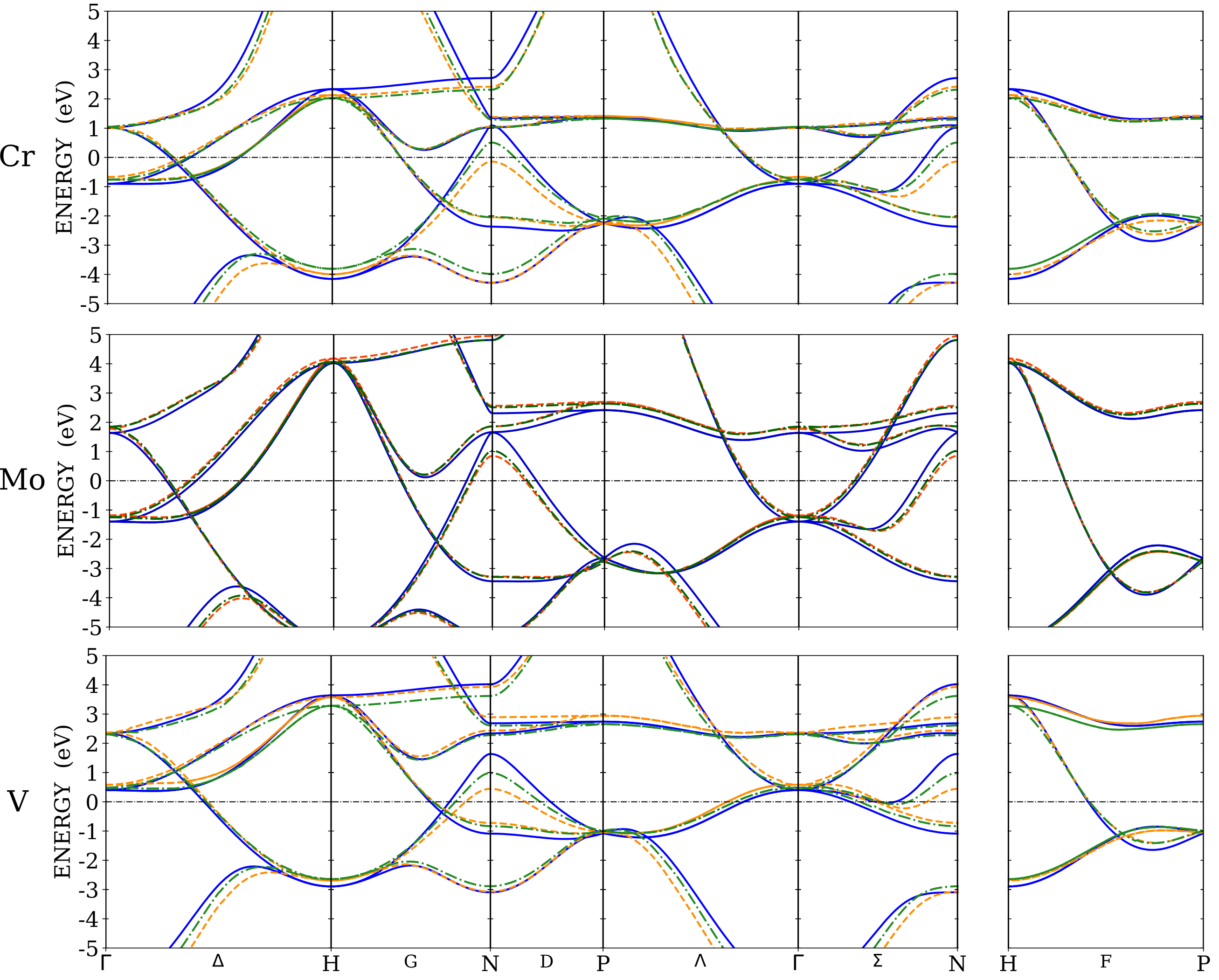}
	\caption{\label{fig:CrMoV_bandstr_QSGW} The DFT (LDA \cite{perdew_accurate_1992}) bandstructure (blue, solid line) compared to the quasiparticle-ized single-step G$^{LDA}$W$^{LDA}$ result (orange/red, dashed line), and the fully quasiparticle self-consistent result (green, dash-dotted line). For nonmagnetic Cr (top) and Mo (middle) and V (bottom). The Fermi energy is at 0.}
\end{figure*}
The results of single-step ($G^{LDA}W^{LDA}$) and QSGW bandstructure calculations for nonmagnetic Cr, Mo, and V are shown in Fig.~\ref{fig:CrMoV_bandstr_QSGW}.
For both the $G^{LDA}W^{LDA}$ and the QSGW calculations the noninteracting, Hermitian approximation to the $GW$ self-energy \cite{vanschilfgaarde_quasiparticle_2006} has been used to obtain the bandstructure.
The most obvious difference between the different $GW$ calculations is to the band that has $p$ character as it crosses the Fermi energy near to the $N$ point. 
For the $G^{LDA}W^{LDA}$ Cr calculation this band never rises above the Fermi energy, meaning that the $N$ hole ellipsoids are completely absent.
In the QSGW calculation, however, this band does cross the Fermi energy, thus the $N$ hole ellipsoids survive, but with a reduced size compared to the DFT-LDA prediction.

For each of the calculations the energy eigenvalue at the $N$ point for the key band that is responsible for the $N$ hole ellipsoids is listed in Table~\ref{tab:N-energy}.
This data gives a sense of the level of dependence of this band on the treatment of the exchange-correlation potential in each of the different metals.
\begin{table}
	\caption{\label{tab:N-energy} (All units are eV, and all values are relative to the Fermi energy at zero.) For each of the calculations, the energy eigenvalue at the $N$ point is listed for the key band that causes the $N$ hole ellipsoids.}
	\begin{ruledtabular}
		\begin{tabular}{c|cccc}
		    Approximation&Cr&Mo&V
		    \\
		    \hline
			Zero\footnote{A DFT calculation where there is zero exchange-correlation potential.}&-0.87&0.70&0.57
			\\
			LDA&1.04&1.62&1.65
			\\
			PBE&0.94&1.60&1.53
			\\
			SCAN&0.61&1.37&1.30
			\\
			LDA\footnote{Questaal LDA calculation.}&1.09&1.65&1.64
			\\
			$G^{LDA}W^{LDA}$&-0.14&0.83&0.44
			\\
			QSGW&0.51&1.02&0.99
			\\
		\end{tabular}
	\end{ruledtabular}
\end{table}

For Cr, the positions where the band crosses the Fermi energy in the QSGW calculations in Fig.~\ref{fig:CrMoV_bandstr_QSGW} and the SCAN calculations in Fig.~\ref{fig:CrMoV_bandstr_SCAN} are quite similar, whereas QSGW affects Mo and V more strongly than SCAN does.
The difference between the $G^{LDA}W^{LDA}$ and QSGW calculations is small for Mo, which still exhibits $N$ hole ellipsoids for $G^{LDA}W^{LDA}$ calculations (a similar result can be obtained for Nb).
We attribute this to the lower overall sensitivity to the exchange-correlation potential in the case of Mo (it is also less sensitive in Nb than it is V, and in general the exchange-correlation potential is expected to be less important to the electronic structure of the fifth period transition metals than it is to the fourth period metals that are in the same group because of the spatial extent of the outermost wave functions). 
In fact, we found that Mo retains all of its Fermi surface features even in the extreme case when the exchange-correlation potential is completely absent from the groundstate DFT calculation. 
In contrast, the existence of the $N$ hole ellipsoids in Cr depends on the inclusion of an exchange-correlation potential (we found that for Cr the $N$ hole ellipsoids vanish if no exchange-correlation potential is included in the calculation, whereas for Mo they do not).

For Cr and Mo any reduction of the $N$ hole ellipsoid sizes that occurs because of changes to the exchange-correlation potential is compensated by a reduction in electron ball volume, so that the average electron occupation across the Brillouin zone is conserved (a hole-like volume has a lower electron occupation density than the average across the whole zone, and an electron-like volume has a higher electron occupation density than the average over the whole zone).
The sizes and shapes of the $\Gamma$ and $H$ octahedra, which are already well predicted by LDA, are rather stable under various exchange-correlation approximations.
(The nesting vector between these features for the LDA calculation is already a good match to the spin-density wave length, and it is therefore reassuring that large changes to this nesting vector do not occur, although it is also worth noting that the requirement that the nesting vector must match the spin-density wave vector has recently been contested \cite{lamago_measurement_2010}.)
In contrast, for V (and Nb) the reduction in $N$ hole ellipsoid volume is compensated by an increase in volume of the other hole-like Fermi surface features.

\subsubsection{Magnetism}
\label{sec:bandstr_mag}

A study of the magnetic state, and the role it might play in altering the Fermi surface, is most important for Cr because it exhibits more complex magnetic order than any of the surrounding elemental transition metals, forming an incommensurate antiferromagnetic SDW at low temperature \cite{fawcett_spin-density-wave_1988}.
The relatively long wavelength of this SDW means that the unit cell is relatively large (approximately 21 simple cubic cells parallel to the SDW vector).
As a result, we found the computational expense of the SDW state too great to rigorously test it with QSGW calculations.
We were able to converge SDW calculations for SCAN, but unfortunately the predicted SDW has a modulation that is closer to square than sinusoidal (SCAN is known to be problematic for describing itinerant magnetism \cite{isaacs_performance_2018,ekholm_assessing_2018,fu_applicability_2018,fu_density_2019}, and also see Sec.~\ref{sec:other}).
The SDW is not found to be the theoretical groundstate in any of these DFT calculations, unless the magnetic moment of each atom is constrained to a particular direction (if unconstrained, the theoretical groundstate is the commensurate antiferromagnetic one).
We therefore mainly test the effects of magnetism here by studying the less computationally expensive commensurate antiferromagnetic state.

The Fermi surface of the nonmagnetic and commensurate antiferromagnetic states of Cr can be compared in the reciprocal-space cell of the 2 atom basis simple cubic (SC) lattice (which is the Brillouin zone for the commensurate antiferromagnetic structure).
The BCC Fermi surface features (Fig.~\ref{fig:CrMoW_FS}) can be connected to the Fermi surface features of the SC zone: the BCC $\Gamma$ and $H$ points both translate to the SC $\Gamma$ point, so the two octahedra (one electron-like and one hole-like) mostly cancel each other in the SC zone; the BCC $N$ point translates to the SC $M$ point, which means that the $N$ hole ellipsoids (BCC) can be related directly to $M$ hole pockets in the SC lattice Brillouin zone; the electron balls can be traced quite directly to the region surrounding the $X$ point of the SC Brillouin zone.

\begin{figure}
	\includegraphics[width=0.45\textwidth]{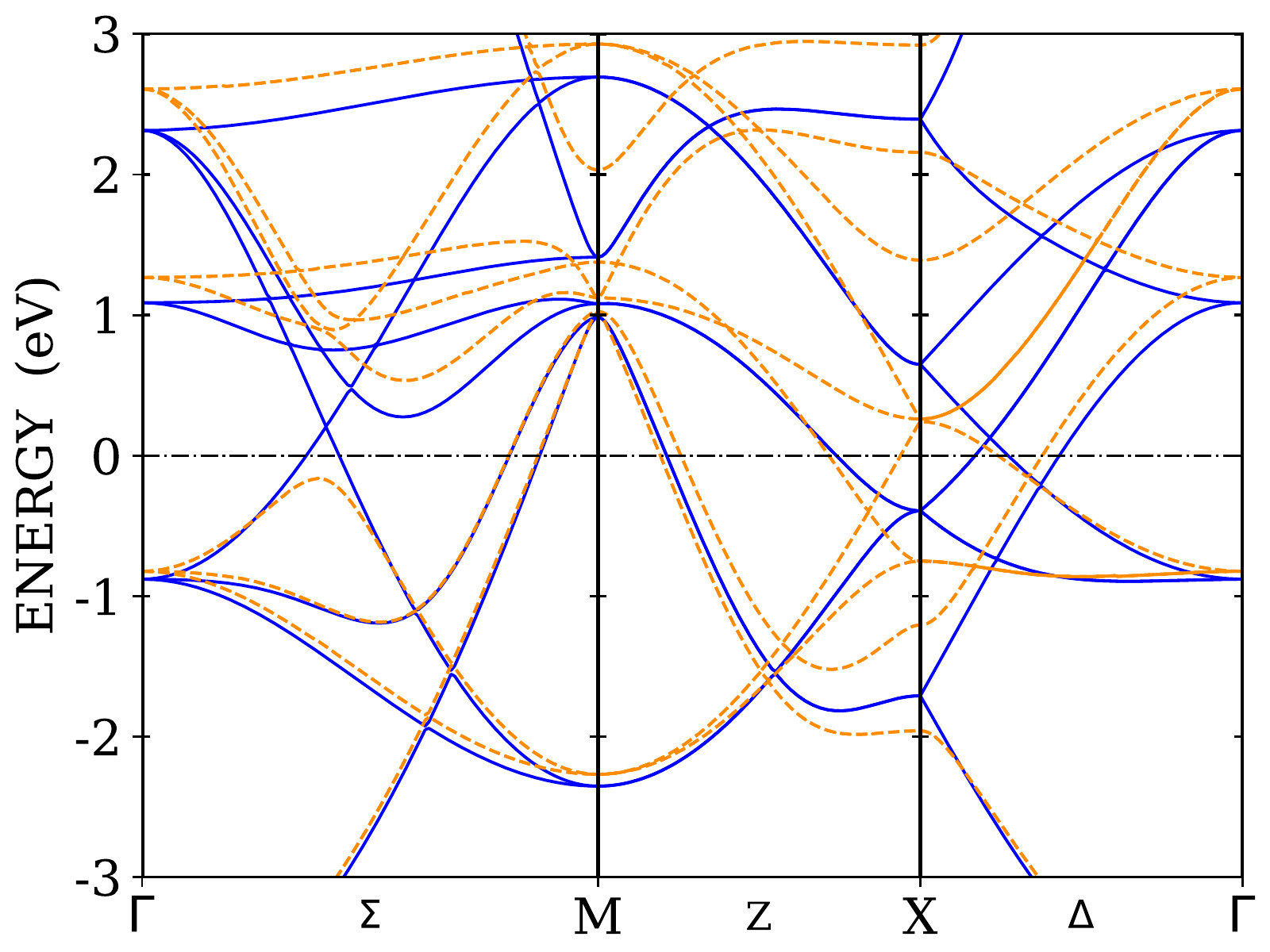}
	\caption{\label{fig:Cr_NM-vs-AFM} The PBE GGA \cite{perdew_generalized_1996} Cr bandstructure for nonmagnetic calculations (blue, solid line) compared to commensurate antiferromagnetic (orange, dashed line), in the Brillouin zone of the simple cubic unit cell. The Fermi energy is at 0.}
\end{figure}

A comparison between the nonmagnetic and commensurate antiferromagnetic bandstructures for a PBE GGA calculation is shown in Figure~\ref{fig:Cr_NM-vs-AFM}.
For this calculation, and others, the magnetic and nonmagnetic energy eigenvalues at the $X$ point are listed in Table~\ref{tab:X_energies}.
LDA, PBE, SCAN, and QSGW all demonstrate similar relationships between magnetic and nonmagnetic calculations (though the size of the magnetic splitting varies).
Our results are consistent with prior comparisons between nonmagnetic and commensurate antiferromagnetic states \cite{kubler_spin-density_1980}. 
\begin{table}
	\caption{\label{tab:X_energies} (All units are eV, and relative to a Fermi energy of 0.) Energy eigenvalues at the $X$ point of the SC Brillouin zone in the $\pm~3$~eV energy window, for nonmagnetic (NM) and commensurate antiferromagnetic (CAFM) calculations.}
	\begin{ruledtabular}
		\begin{tabular}{cccccccc}
		    \multicolumn{2}{c}{LDA \cite{perdew_accurate_1992}}&\multicolumn{2}{c}{PBE \cite{perdew_generalized_1996}}&\multicolumn{2}{c}{SCAN \cite{sun_strongly_2015}}&\multicolumn{2}{c}{QSGW \footnote{A Questaal \cite{pashov_questaal_2020} calculation (the other values in this table are from Elk \cite{noauthor_elk_nodate}).}}
		    \\
		    NM&CAFM&NM&CAFM&NM&CAFM&NM&CAFM
		    \\
		    \hline
			-1.75&-1.87&-1.71&-1.96&-1.58&-2.26&-1.57&-1.89
			\\
            -1.75&-1.55&-1.71&-1.21&-1.58&-1.50&-1.57&-0.76
            \\
            -0.38&-0.54&-0.39&-0.75&-0.47&-1.49&-0.27&-0.76
            \\
            -0.38&-0.54&-0.39&-0.75&-0.47&-0.66&-0.27&-0.73
            \\
            -0.38&-0.14&-0.39&0.24&-0.47&-0.47&-0.27&0.09
            \\
            -0.38&-0.14&-0.39&0.26&-0.47&0.93&-0.27&0.77
            \\
            0.66&0.48&0.65&0.26&0.67&0.93&0.70&0.77
            \\
            0.66&0.94&0.65&1.39&0.67&2.10&0.70&1.76
            \\
            2.36&2.23&2.39&2.16&2.72&2.31&2.12&1.93
            \\
            2.36&2.56&2.39&2.92&2.76&3.81&2.12&3.23
			\\
		\end{tabular}
	\end{ruledtabular}
\end{table}

There are two notable Fermi surface differences between these nonmagnetic and magnetic calculations. 
The first is that the nonmagnetic bandstructure has an additional hole-like Fermi surface feature (this can be seen along the ${\Gamma}M$ high symmetry line in Fig.~\ref{fig:Cr_NM-vs-AFM}). 
This feature arises because the $H$ hole octahedron (BCC) is slightly larger than the $\Gamma$ electron octahedron (BCC), so that when both are folded to the $\Gamma$ point of the SC lattice Brillouin zone there is a small region where these two octahedra do not counterbalance each other.
In the commensurate antiferromagnetic bandstructure this hole-like Fermi surface feature becomes gapped.
The second notable difference is that the nonmagnetic state has higher occupation at the $X$ point of the SC Brillouin zone, where there are fewer bands beneath the Fermi energy in the commensurate antiferromagnetic calculation.
These two differences between the magnetic and nonmagnetic calculations compensate each other. 
There is a hole-like Fermi surface that arises because the $H$ hole octahedra are slightly larger than the $\Gamma$ electron octahedra, and when this feature becomes magnetically gapped and vanishes some hole volume is lost.
To compensate this loss of hole volume (and conserve total electron occupation number), electron-like Fermi surface volume is lost around the $X$ point.

Despite the fact that overall bandstructure and Fermi surface changes do occur between the nonmagnetic and commensurate antiferromagnetic calculations, there is very little difference between the size of the $M$ hole pockets (along $M{\Gamma}$ there is no difference, and there are only very small differences along $MX$ and $MR$).
The fact that this feature does not depend much on the magnetic state is somewhat predictable from its $p$ character (Fig.~\ref{fig:Cr_bandstr_lspec}).

Returning to the BCC lattice Brillouin zone Fermi surface, we therefore find that differences between experiment and theory over the size of the $N$ hole ellipsoids cannot be explained by their magnetic states being different. 
The sizes of the electron balls, however, do depend on the magnetic state.
These reduce in size when the $\Gamma$ and $H$ octahedra become magnetically gapped, conserving the electron occupied volume in the Brillouin zone.

\subsection{dHvA data}
\label{sec:dHvA}


\begin{table*}
	\caption{\label{tab:CrdHvA}(All units are \r{A}$^{-1}$.) Table of Fermi surface caliper data for BCC Cr. Calculations are nonmagnetic, and are compared to values derived from experimental dHvA measurements of the antiferromagnetic spin-density wave state of Cr \cite{graebner_haas-van_1968}.}
	\begin{ruledtabular}
		\begin{tabular}{cc|ccccc|c}
			&&\multicolumn{3}{c}{DFT \cite{noauthor_elk_nodate}}&\multicolumn{2}{c}{$GW$ \cite{pashov_questaal_2020}}&Experiment
			\\
		    Feature&&LDA \cite{perdew_accurate_1992}&PBE \cite{perdew_generalized_1996}&SCAN \cite{sun_strongly_2015}&$G^{LDA}W^{LDA}$&QSGW&dHvA \cite{graebner_haas-van_1968}
			\\ \hline
			$N$ hole ellipsoid&$NP$&0.32&0.30&0.24&-\footnote{The $N$ hole ellipsoid is not present in the $G^{LDA}W^{LDA}$ calculation.}&0.23&0.27
			\\
			&$N\Gamma$&0.31&0.29&0.24&-\footnotemark[1]&0.22&0.23
			\\
			&$NH$&0.20&0.20&0.18&-\footnotemark[1]&0.15&0.17
			\\
			\hline
			$H$ hole octahedron&$H\Gamma$&0.90&0.90&0.87&0.92&0.92&-\footnote{The $H$ and $\Gamma$ octahedra are gapped in the presence of the spin-density wave and therefore do not appear in the experimental dHvA measurements.}
			\\
			&$HP$&0.56&0.56&0.55&0.56&0.56&-\footnotemark[2]
			\\
			&$HN$&0.66&0.66&0.64&0.66&0.67&-\footnotemark[2]
			\\ 
			\hline
			$\Gamma$ electron octahedron&${\Gamma}H$&1.28&1.28&1.31&1.26&1.26&-\footnotemark[2]
			\\
			&${\Gamma}P$&0.51&0.51&0.51&0.48&0.49&-\footnotemark[2]
			\\
			&${\Gamma}N$&0.56&0.56&0.57&0.53&0.53&-\footnotemark[2]
			\\
			\hline 
			Electron ball&${\Gamma}H$&0.69&0.67&0.61&0.61&0.62&0.51
			\\
			&$\perp {\Gamma}H$  \footnote{Measured in the ${\Gamma}NH$ plane (see Fig.~\ref{fig:Cr_FS_plane}), along the $N$-$N$ line.}&0.60&0.60&0.56&0.52&0.56&0.50
			\\	
		\end{tabular}
	\end{ruledtabular}
\end{table*}

\begin{table*}
	\caption{\label{tab:Mo-dHvA} (All units are \r{A}$^{-1}$.) Table of Fermi surface caliper data for Mo. Calculations are nonmagnetic, compared to values derived from experimental dHvA measurements of paramagnetic Mo.}
	\begin{ruledtabular}
		\begin{tabular}{cc|ccccc|ccc}
			&&\multicolumn{3}{c}{DFT \cite{noauthor_elk_nodate}}&\multicolumn{2}{c}{$GW$ \cite{pashov_questaal_2020}}&\multicolumn{3}{c}{Experiment}
			\\
			Feature&&LDA \cite{perdew_accurate_1992}&PBE \cite{perdew_generalized_1996}&SCAN \cite{sun_strongly_2015}&$G^{LDA}W^{LDA}$&QSGW&dHvA \cite{sparlin_empirical_1966}&dHvA \cite{hoekstra_determination_1973}&RFSE\footnote{Radio frequency size effect measurements.} \cite{boiko_investigation_1969}
			\\ \hline
			$N$ hole ellipsoid&$NP$&0.38&0.39&0.35&0.29&0.31&0.39&0.37&0.38
			\\
			&$N{\Gamma}$&0.35&0.35&0.31&0.25&0.28&0.30&0.33&0.29
			\\
			&$NH$&0.23&0.23&0.22&0.18&0.19&0.23&0.22&0.22
			\\
			\hline
			$H$ hole octahedron&$H{\Gamma}$&0.82&0.82&0.82&0.83&0.83&-\footnote{The authors of Ref.~\cite{sparlin_empirical_1966} were not able to observe enough oscillations corresponding to the $\Gamma$ or $H$ octahedra to estimate these values.}&0.81&0.79
			\\
			&$HP$&0.50&0.50&0.50&0.51&0.50&-\footnotemark[2]&0.49&0.51
			\\
			&$HN$&0.59&0.60&0.60&0.61&0.60&-\footnotemark[2]&0.61&0.60
			\\ 
			\hline
			$\Gamma$ electron octahedron&${\Gamma}H$&1.18&1.18&1.18&1.17&1.18&-\footnotemark[2]&-\footnote{Although oscillations were observed, the authors of Ref.~\cite{hoekstra_determination_1973} argue that it is not possible to quote an accurate value from their measurements, nor from the measurements in Ref.~\cite{boiko_investigation_1969}.}&-\footnotemark[3]
			\\
			&${\Gamma}P$&0.48&0.47&0.47&0.44&0.45&-\footnotemark[2]&-\footnotemark[3]&-\footnotemark[3]
			\\
			&${\Gamma}N$&0.53&0.53&0.51&0.49&0.50&-\footnotemark[2]&-\footnotemark[3]&-\footnotemark[3]
			\\
			\hline
			Electron ball&${\Gamma}H$&0.69&0.67&0.63&0.64&0.64&0.73\footnote{This was estimated by the authors of Ref.~\cite{sparlin_empirical_1966} using a model Fermi surface shape that is probably too crude, yielding an inaccurate value. }&-\footnotemark[3]&-\footnotemark[3]
			\\
			&$\perp {\Gamma}H$ \footnote{Measured in the ${\Gamma}NH$ plane (see Fig.~\ref{fig:Cr_FS_plane}), along the $N$-$N$ line.} &0.59&0.59&0.56&0.52&0.53&0.53&-\footnotemark[3]&-\footnotemark[3]
			\\
			\hline
			Electron lens&${\Gamma}H$&0.28&0.26&0.20&0.17&0.19&-\footnote{A value is not quoted by the authors of Ref.~\cite{sparlin_empirical_1966}.}&0.22&0.22
			\\
			&$\perp {\Gamma}H$  \footnotemark[5]&0.39&0.39&0.35&0.29&0.30&0.32&0.32&0.31
			\\
		\end{tabular}
	\end{ruledtabular}
\end{table*}

\begin{table*}
    \caption{\label{tab:Mo-dHvA-areas} (All units are \r{A}$^{-2}$.) Table of Fermi surface extremal areas for the $N$ hole ellipsoids of Mo. Calculations are nonmagnetic, compared to experimental measurements on paramagnetic Mo.}
    \begin{ruledtabular}
        \begin{tabular}{c|ccccc|cc}
            &\multicolumn{3}{c}{DFT \cite{noauthor_elk_nodate}}&\multicolumn{2}{c}{$GW$ \cite{pashov_questaal_2020}}&\multicolumn{2}{c}{Experiment}
			\\
            Field dir.&LDA \cite{perdew_accurate_1992}&PBE \cite{perdew_generalized_1996}&SCAN \cite{sun_strongly_2015}&$G^{LDA}W^{LDA}$&QSGW&dHvA \cite{sparlin_empirical_1966}&dHvA \cite{hoekstra_determination_1973}
            \\
            \hline
            {[100]}&0.238&0.235&0.219&0.137&0.162&0.220&0.217
            \\
            &0.320&0.312&0.286&0.186&0.215&0.297&0.285
            \\
            {[110]}&0.265&0.260&0.242&0.155&0.181&0.246&0.240
            \\
            &0.281&0.274&0.257&0.166&0.191&0.282&0.251
            \\
            &0.395&0.385&0.334&0.218&0.256&0.368&0.347
        \end{tabular}
    \end{ruledtabular}
\end{table*}

\begin{table*}
	\caption{\label{tab:V-dHvA} (All units are \r{A}$^{-1}$.) Table of Fermi surface caliper data for V. Calculations are nonmagnetic, compared to values derived from experimental dHvA measurements of paramagnetic V.}
	\begin{ruledtabular}
		\begin{tabular}{cc|ccccc|cc}
			&&\multicolumn{3}{c}{DFT \cite{noauthor_elk_nodate}}&\multicolumn{2}{c}{$GW$ \cite{pashov_questaal_2020}}&\multicolumn{2}{c}{Experiment}
			\\
			Feature&&LDA \cite{perdew_accurate_1992}&PBE \cite{perdew_generalized_1996}&SCAN \cite{sun_strongly_2015}&G$^{LDA}$W$^{LDA}$&QSGW&dHvA \cite{phillips_haas-van_1971}&MTO\footnote{Magnetothermal oscillations} \cite{parker_experimental_1974}
			\\ \hline
			$N$ hole ellipsoid&$NP$&0.56&0.54&0.50&0.32&0.46&0.44&0.46
			\\
			&$N{\Gamma}$&0.59&0.56&0.51&0.28&0.45&0.42&0.44
			\\
			&$NH$&0.38&0.37&0.36&0.23&0.33&0.35&0.36
			\\
		\end{tabular}
	\end{ruledtabular}
\end{table*}

\begin{table*}
    \caption{\label{tab:V-dHvA-areas} (All units are \r{A}$^{-2}$.) Table of Fermi surface extremal areas for the $N$ hole ellipsoids of V. Calcualations are nonmagnetic, compared to experimental measurements on paramagnetic V.}
    \begin{ruledtabular}
        \begin{tabular}{c|ccccc|cc}
            &\multicolumn{3}{c}{DFT \cite{noauthor_elk_nodate}}&\multicolumn{2}{c}{$GW$ \cite{pashov_questaal_2020}}&\multicolumn{2}{c}{Experiment}
			\\
            Field dir.&LDA \cite{perdew_accurate_1992}&PBE \cite{perdew_generalized_1996}&SCAN \cite{sun_strongly_2015}&$G^{LDA}W^{LDA}$&QSGW&dHvA \cite{phillips_haas-van_1971}&MTO\footnote{Magnetothermal oscillations}\cite{parker_experimental_1974}
            \\
            \hline
            {[100]}&0.658&0.620&0.552&0.202&0.449&0.502&0.505
            \\
            &0.746&0.701&0.617&0.250&0.514&0.575&0.576
            \\
            {[110]}&0.644&0.610&0.546&0.219&0.454&0.504&0.504
            \\
            &0.689&0.649&0.581&0.236&0.476&0.532&0.531
            \\
            &0.822&0.773&0.665&0.270&0.566&0.641&0.641
        \end{tabular}
    \end{ruledtabular}
\end{table*}

The sizes (calipers) of the Cr Fermi surface features along high symmetry lines for various nonmagnetic calculations and for antiferromagnetic SDW state dHvA measurements are documented in Table~\ref{tab:CrdHvA}. 
The calipers from our LDA calculation are mostly consistent with the previous ones by Laurent \textit{et al.} \cite{laurent_band_1981}, except for the size of the electron ball. 
Because of this, we cross-checked our DFT-LDA calculations using different modern codes that employ different basis types (Elk \cite{noauthor_elk_nodate} and Questaal \cite{pashov_questaal_2020}), and found these predicted essentially identical Fermi surfaces (to each other) when the same exchange-correlation functional is used.
(For the LDA the energy difference between respective band energies for the different codes is 0.05~eV or smaller.)
Other LDA calculations also show an electron ball size more consistent with that which we obtained \cite{kubler_spin-density_1980}.
Newer DFT codes typically demonstrate a higher level of precision than older ones  \cite{lejaeghere_reproducibility_2016} (and precision is expected to be especially high for full potential calculations, such as those employed here.)
We therefore find that the experiment-theory discrepancy is worse than previously noted: both the electron balls and the $N$ hole ellipsoids from the dHvA data are poorly predicted by the LDA/GGA, not only the $N$ hole ellipsoids.
It makes sense that if the $N$ hole ellipsoids are predicted to be larger than the experimental ones, then an electron-like Fermi surface sheet must also be too large.

For Cr the Fermi surfaces for DFT calculations with the LDA \cite{perdew_accurate_1992} and PBE GGA \cite{perdew_generalized_1996} are very similar (as are the bandstructures).
The SCAN and QSGW calculations produce $N$ hole ellipsoid sizes that are closer to the measured dHvA ones than LDA/GGA calculations do.
The size of the electron balls is also improved, though these are still notably too large compared to the experimental values; this is the largest remaining discrepancy with the dHvA data for SCAN for any of the metals that we tested.
However, we found that the sizes of the electron balls can be reduced in magnetic calculations (Sec.~\ref{sec:bandstr_mag}).
It therefore seems reasonable to attribute most of the remaining discrepancy to the difference between the nonmagnetic (calculated) and incommensurate antiferromagnetic SDW (measured) Fermi surfaces.

The calipers of the Mo Fermi surface features along high symmetry lines for various nonmagnetic calculations and paramagnetic dHvA measurements are documented in Table~\ref{tab:Mo-dHvA}. 
Direct comparisons to the measured extremal areas are documented in Table~\ref{tab:Mo-dHvA-areas}.
The $G^{LDA}W^{LDA}$ and QSGW methods produce Fermi surfaces which are more similar to each other for Mo than they are for Cr.
Both of these $GW$ calculations produce $N$ hole ellipsoids that are too small compared to the experimental values (the $NP$ caliper, in particular).
The $N$ hole ellipsoid sizes are also reduced in the SCAN calculations, but not so much as to make the agreement with experimental data any worse than LDA/GGA. In fact, the agreement is a little bit better for SCAN than LDA/GGA.
This is more apparent from the extremal areas that are documented in Table~\ref{tab:Mo-dHvA-areas} than from the comparatively less reliable calipers documented in Table~\ref{tab:Mo-dHvA}  (since these calipers are derived from the areas with the help of a model).
For the extremal areas, SCAN deviates by 4\% at worst from the most recent dHvA measurement \cite{hoekstra_determination_1973}, whereas LDA deviates by up to 14\%.

The Group V transition metals V and Nb also have ellipsoidal $N$ hole pockets that can be compared, and size comparisons for these are listed in Table~\ref{tab:V-dHvA}.
But it should be noted that the exact shapes of the $N$ hole ellipsoids in V seem to deviate a little more from conventional ellipsoids than they do for Cr and Mo, and this means the experimentally derived calipers listed in Table~\ref{tab:V-dHvA} are more likely to be erroneous.
It is therefore important to compare directly to the measured extremal areas, which are listed in Table~\ref{tab:V-dHvA-areas}.
Considering these extremal area values, SCAN deviates by 9\% at most, whereas LDA deviates by up to 36\%.
The trend of these results is similar to that obtained for Cr and Mo; the size of the $N$ hole ellipsoids is reduced in SCAN and QSGW calculations.
Note that SCAN and $GW$ calculations on V also show an increase in volume of the other hole-like Fermi surface features (which compensates the reduction in volume of the $N$ hole ellipsoids) and this is consistent with the experimental data too \cite{parker_experimental_1974}.
We also find that the exchange-correlation approximation has a much smaller effect on the bandstructure of Nb than it does for V, which is consistent with the relationship between Cr and Mo, and consistent with the expectation that fifth period transition metals should be less correlated than the corresponding fourth period ones.

\begin{figure}
	\includegraphics[width=0.49\textwidth]{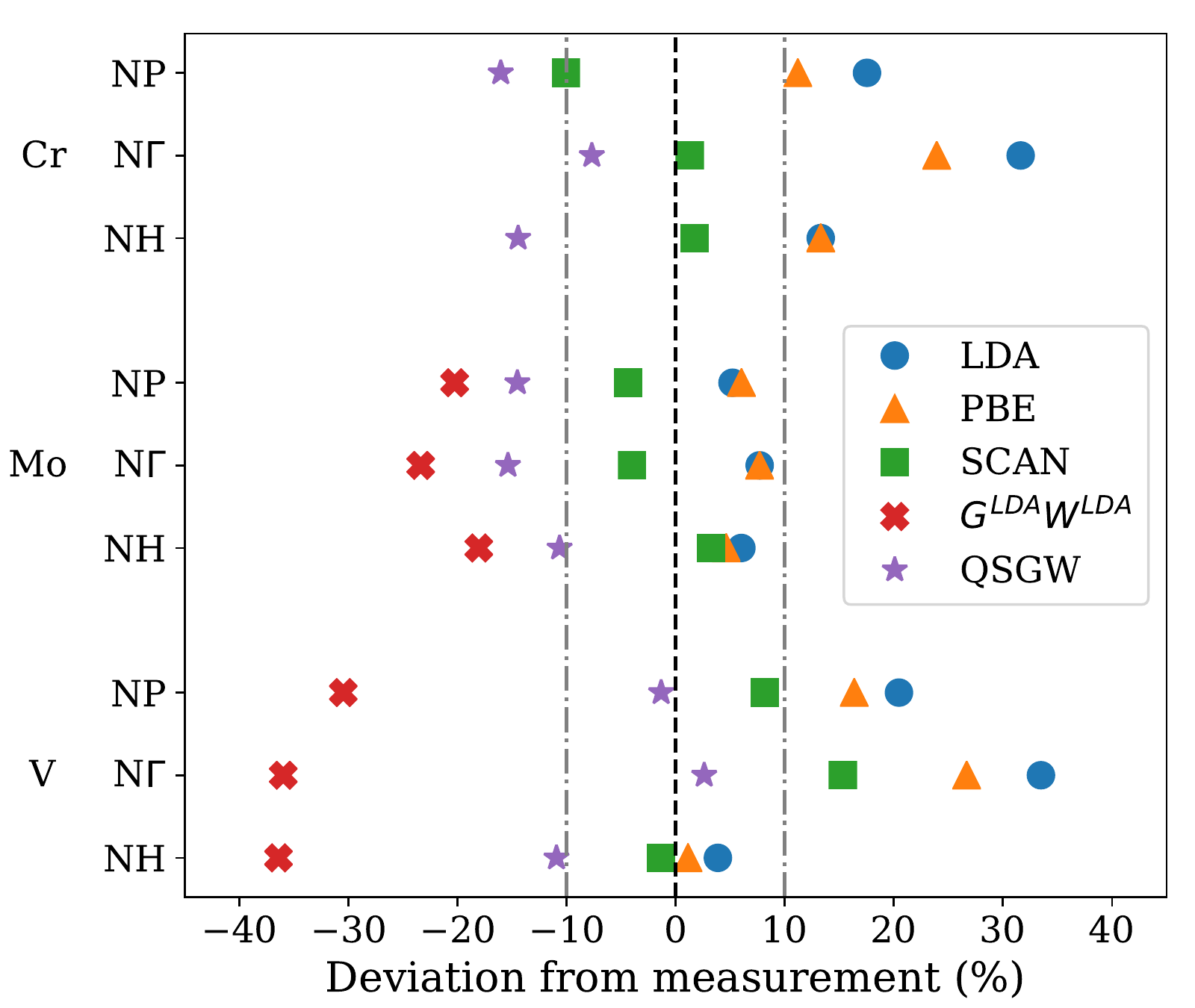}
	\caption{\label{fig:N_sizes} }
	Comparison of calculated $N$ hole ellipsoid size calipers to values derived from dHvA experimental measurements (\cite{graebner_haas-van_1968,hoekstra_determination_1973,phillips_haas-van_1971}). Calculations are nonmagnetic, compared to experimental measurements on the magnetic state (which is the SDW for Cr, and paramagmetic for Mo and V). Vertical lines representing 0\% and 10\% error are included as a visual guide. The values are listed in Tables~\ref{tab:CrdHvA}, \ref{tab:Mo-dHvA} and  \ref{tab:V-dHvA}.
\end{figure}
\begin{figure}
	\includegraphics[width=0.49\textwidth]{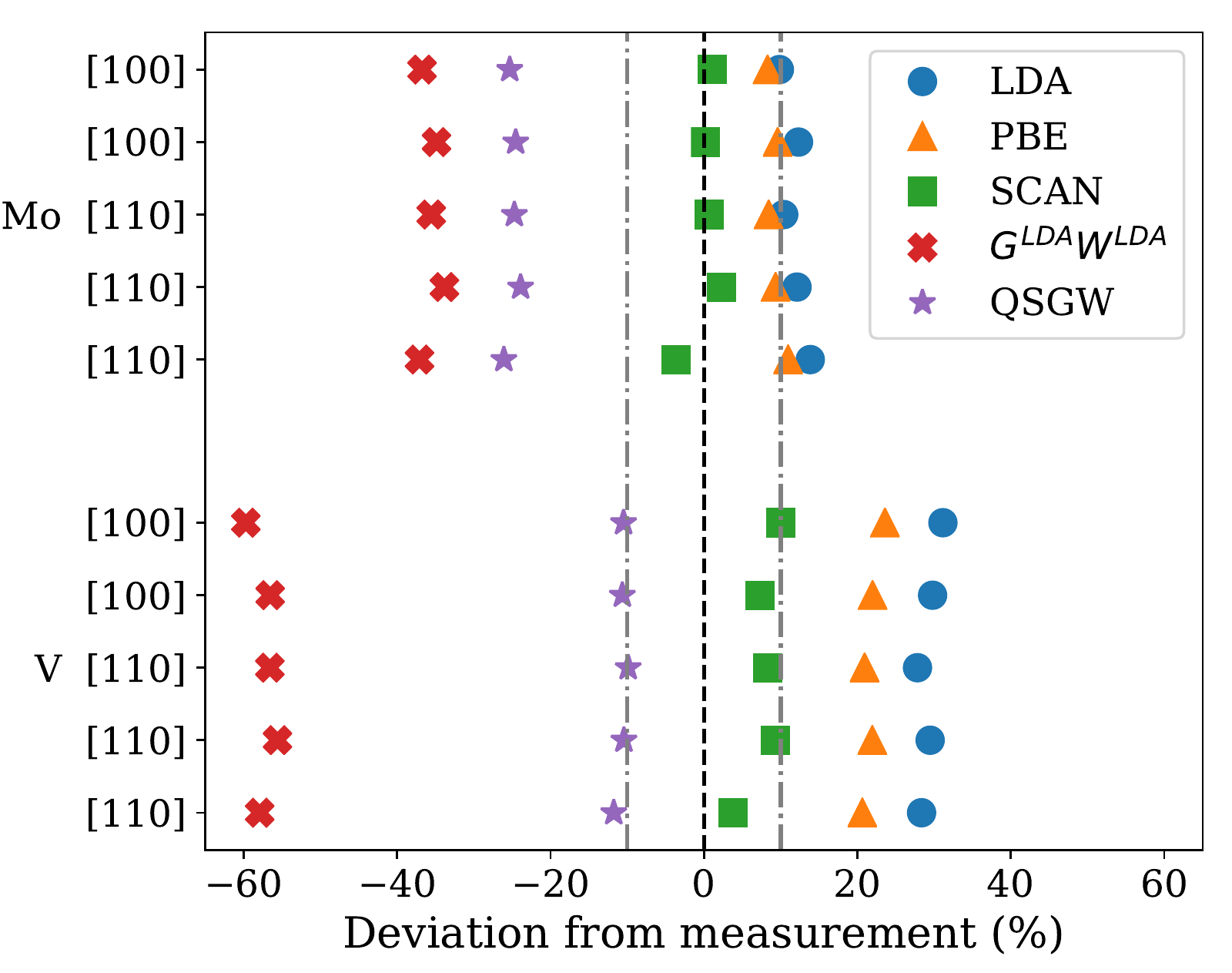}
	\caption{\label{fig:N_sizes_areas} }
	Comparison of calculated $N$ hole ellipsoid extremal areas to dHvA experimental measurements (\cite{hoekstra_determination_1973,phillips_haas-van_1971}). Calculations are nonmagnetic, compared to experimental measurements on the paramagnetic state. Vertical lines representing 0\% and 10\% error are included as a visual guide. The values are listed in Tables~\ref{tab:Mo-dHvA-areas}, and \ref{tab:V-dHvA-areas}.
\end{figure}
For Cr, V, and Mo, the $N$ hole ellipsoid size calipers are visualized in Fig.~\ref{fig:N_sizes}, and extremal areas are visualized in Fig.~\ref{fig:N_sizes_areas} (because of the SDW and tetragonal symmetry the Cr extremal areas are more complicated, so have been excluded).
Considering all of the elemental metals tested here, and all of the exchange-correlation approximations, it is the predictions of SCAN that are most accurate, only exceeding 10\% deviation from the dHvA experimental caliper for one of the nine sizes listed in Fig.~\ref{fig:N_sizes} (for V along the $N{\Gamma}$ direction).
That the error on this one value is larger than 10\% is not necessarily a cause for concern because the band that controls the size of the V $N$ hole ellipsoid has a low gradient along $N\Gamma$ (see Fig.~\ref{fig:CrMoV_bandstr_QSGW}), and is therefore a lower quality indicator of the accuracy of the theoretical prediction, because smaller changes to the band shape and position could result in larger changes to the Fermi surface.
Along this $N\Gamma$ high symmetry line the Fermi surface for V deviates from the ellipsoidal model that was used to obtain size calipers from the dHvA extremal area data, so it is better to compare directly to the experimentally measured extremal areas instead.
For these extremal areas, the SCAN calculations do not exceed 10\% disagreement.
For Mo the calipers and extremal areas are quite consistent, which is a reflection of the fact that the ellipsoidal model is better for the Fermi surfaces of the Group VI metals.

\subsection{2D-ACAR data}
\label{sec:2D-ACAR}

\begin{figure*}
	\includegraphics[width=0.9\textwidth]{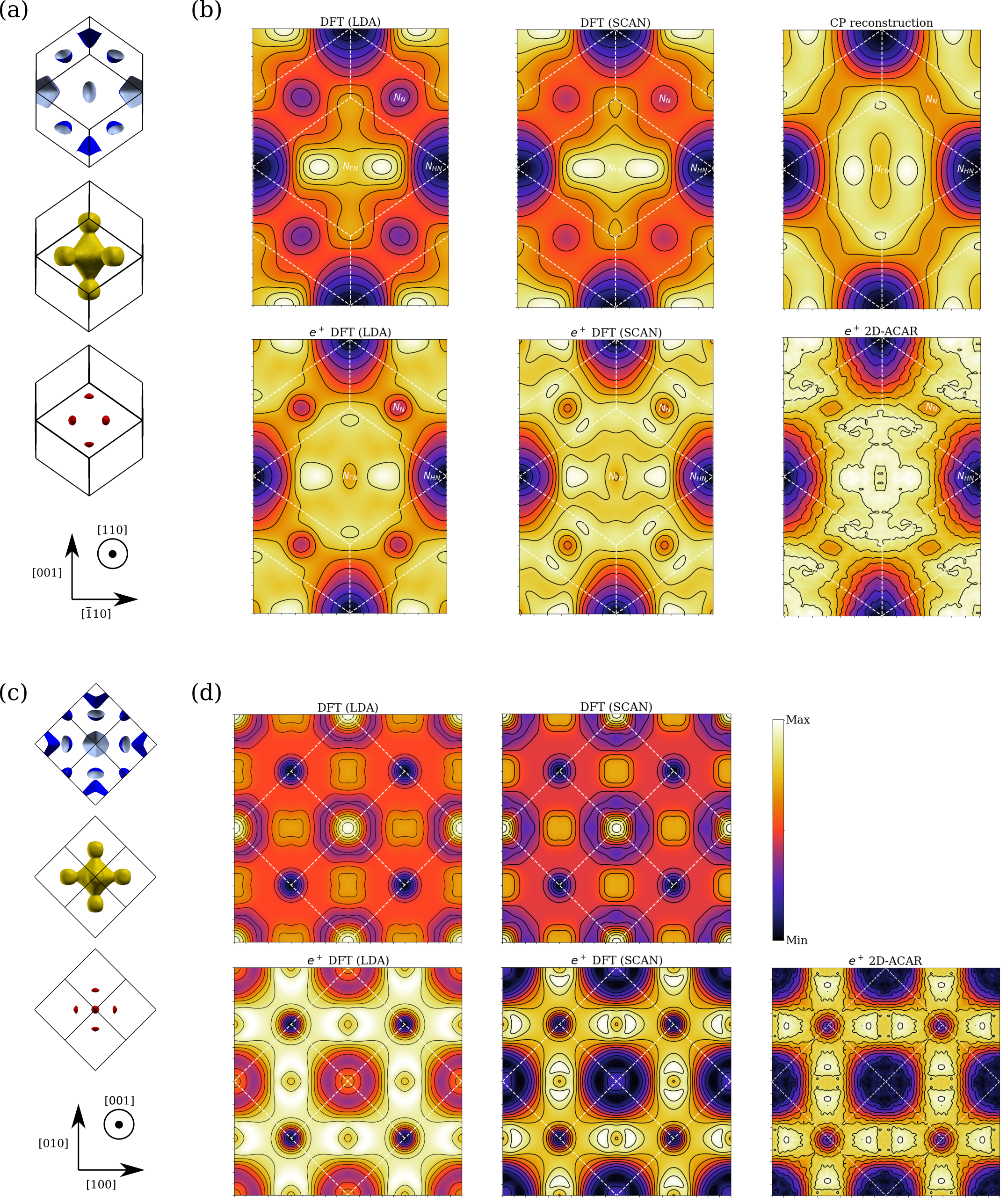}
	\caption{\label{fig:Cr_110_100_EMD} }
	a) and c) The 3D Fermi surface of nonmagnetic Cr according to DFT-LDA calculations, rotated to the match the orientation of the projected 2D occupation densities.
	b) and d) Cr 2D (once integrated) occupation densities for the [110] and [001] projection directions. Various theoretical calculations are compared to experimental results (2D-ACAR \cite{fretwell_reconstruction_1995} and a reconstruction from Compton profile measurements \cite{dugdale_high-resolution_2000,tanaka_study_2000}, which is referred to as `CP reconstruction'). The projection of the Brillouin zone is indicated by the white dashed lines. The theoretical data has been convoluted with an appropriate experimental resolution function. The results that are marked $e^+$ include the influence of the positron (using the enhancement of Drummond \textit{et al.} \cite{drummond_quantum_2011} for theoretical results). 
\end{figure*}

Comparisons between the calculated Cr Fermi surfaces and experimental 2D-ACAR positron annihilation measurements (\cite{fretwell_reconstruction_1995}) are shown in Fig.~\ref{fig:Cr_110_100_EMD} ([110] and [001] projections of the Fermi surface). 
The calculations are presented both with and without inclusion of the perturbations to the electron density due to the electronic screening of the positively charged positron \cite{drummond_quantum_2011} in 2D-ACAR measurements.
It is currently beyond our capability to account for the influence of the positron in $GW$ approximation calculations, so we focus solely on DFT calculations in this subsection.

For the [110] projections a reconstruction of the 2D occupation density from 1D Compton profiles has also been included in Fig.~\ref{fig:Cr_110_100_EMD}.
The Compton profile measurements \cite{dugdale_high-resolution_2000,tanaka_study_2000} were performed to provide corroborating evidence that the real $N$ hole ellipsoids (which are projected onto themselves at the point labelled $N_N$) are different from LDA predictions, since the photons that are used as a probe in Compton scattering do not perturb the electronic ground state (whereas positrons do).
However, this reconstruction from Compton profiles does suffer from lower experimental momentum resolution than the 2D-ACAR measurement.
(The experimental resolution for the Compton profiles from which this 2D projection was reconstructed is approximately Gaussian with a full width at half maximum (FWHM) of 0.18 a.u., whereas for the [110] 2D-ACAR measurements it is anisotropic with a FWHM of 0.11 a.u. along [$\overline{1}$10] and 0.15 a.u. along [001], using the directions indicated in Fig.~\ref{fig:Cr_110_100_EMD}.)

Based on the difference between the theoretical calculation with and without positron enhancement it can be inferred that the influence of the positron in Cr is quite large, although not so large as to make the underlying Fermi surface entirely unrecognisable.
Differences between the LDA and SCAN occupation densities for Cr are increased by inclusion of the theoretically predicted influence of the positrons.
For both of these projection directions ([110] and [100]) the SCAN calculations clearly demonstrate better agreement with the experimental data than LDA calculations.
The QSGW bandstructure calculations are similar to SCAN ones for nonmagnetic Cr, so a similar improvement relative to LDA may be expected.
For the [110] projection, some discrepancies remain between the SCAN calculation and the experiment. 
However, we note that the SCAN result is exceptionally similar to the `maximum entropy filtered' experimental distribution for the [110] projection \cite{dugdale_fermiology_1998-1}.
The agreement between the [100] projection measurement and the SCAN calculation is very good. 
The maximum entropy filtering procedure was thought to help suppress some of the modulations to the occupation density introduced by a k-dependent positron wave function, and it therefore is to be expected that different projections will be impacted to different extents, depending on how constructively these modulations sum \cite{lock_positron_1973,obrien_enhancement_1995}. 

\begin{figure*}
	\includegraphics[width=0.9\textwidth]{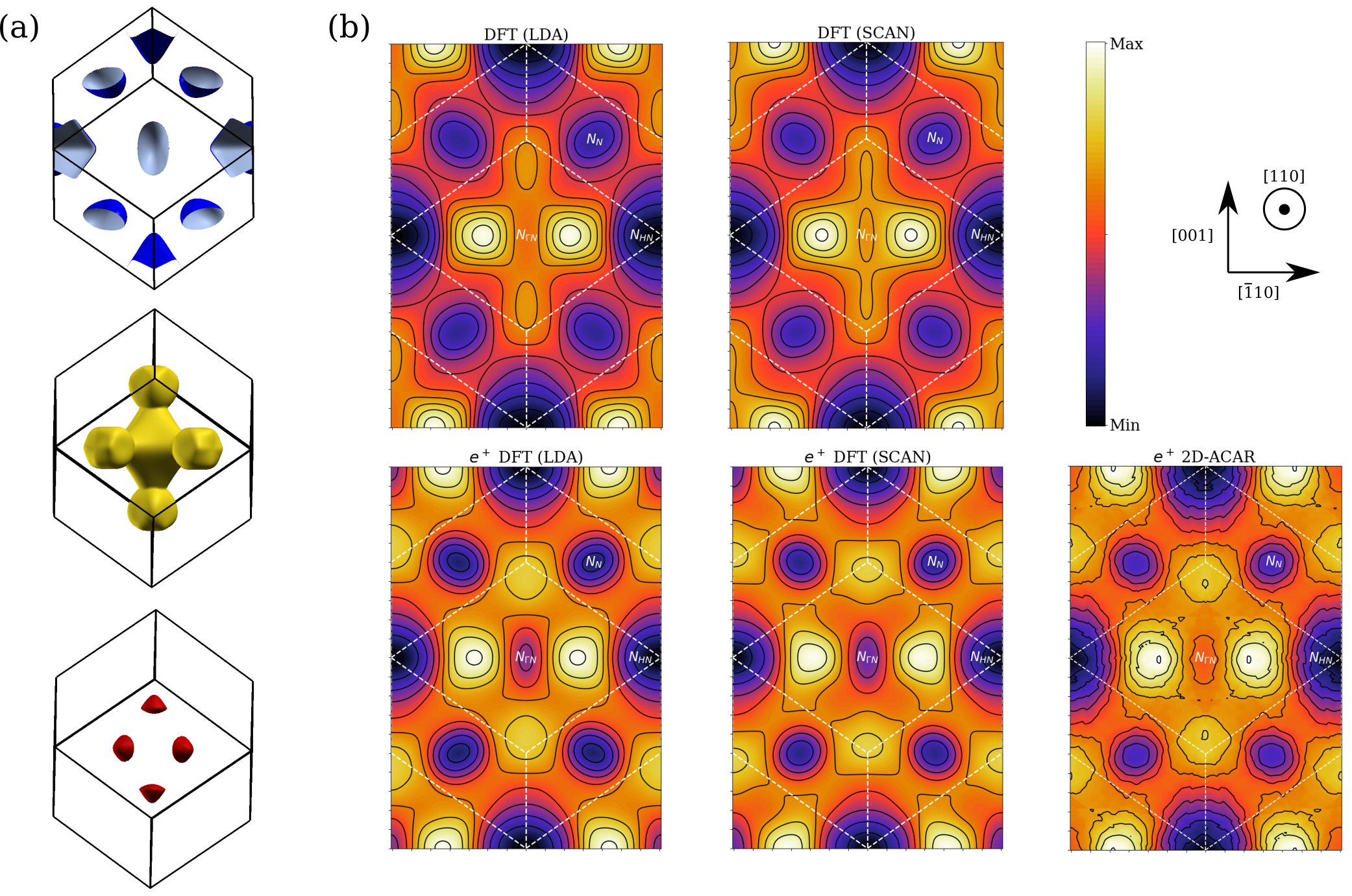}
	\caption{\label{fig:Mo_110_EMD} }
	a) The 3D Fermi surface of nonmagnetic Mo according to DFT-LDA calculations, rotated to the match the orientation of the projected 2D occupation densities.
	b) Mo 2D (once integrated) occupation densities projected along the [110] direction. Experimental results (bottom right)  \cite{hughes_evolution_2004,dugdale_fermiology_1998} are compared to various theoretical calculations. The projection of the BCC Brillouin zone is indicated by the white dashed lines. The theoretical data has been convoluted with an appropriate experimental resolution function. The theoretical occupation densities on the bottom row include the influence of the positron on the measured electronic structure (using the enhancement of Drummond \textit{et al.} \cite{drummond_quantum_2011} for theoretical results), whereas plots on the top row do not.
\end{figure*}
Figure~\ref{fig:Mo_110_EMD} shows how the different theoretical calculations compare to the 2D-ACAR measurements on paramagnetic Mo \cite{hughes_evolution_2004,dugdale_fermiology_1998}, with their projection direction along [110].
It is clear that the influence of the positron in the theoretical calculations is much smaller for Mo than it is for Cr.
Importantly, the LDA and SCAN calculations are more similar than they are for Cr (as might be anticipated from the bandstructures shown in Fig.~\ref{fig:CrMoV_bandstr_SCAN}), and both of these calculations match the experimental data fairly well.
Neither the LDA nor SCAN calculation is in better agreement with the experimental data overall, although there are minor differences.
The SCAN calculation appears to predict the sizes of the $N$ hole pockets more accurately than LDA, whereas LDA is perhaps slightly better in the remainder of the zone.

\begin{figure*}
	\includegraphics[width=0.9\textwidth]{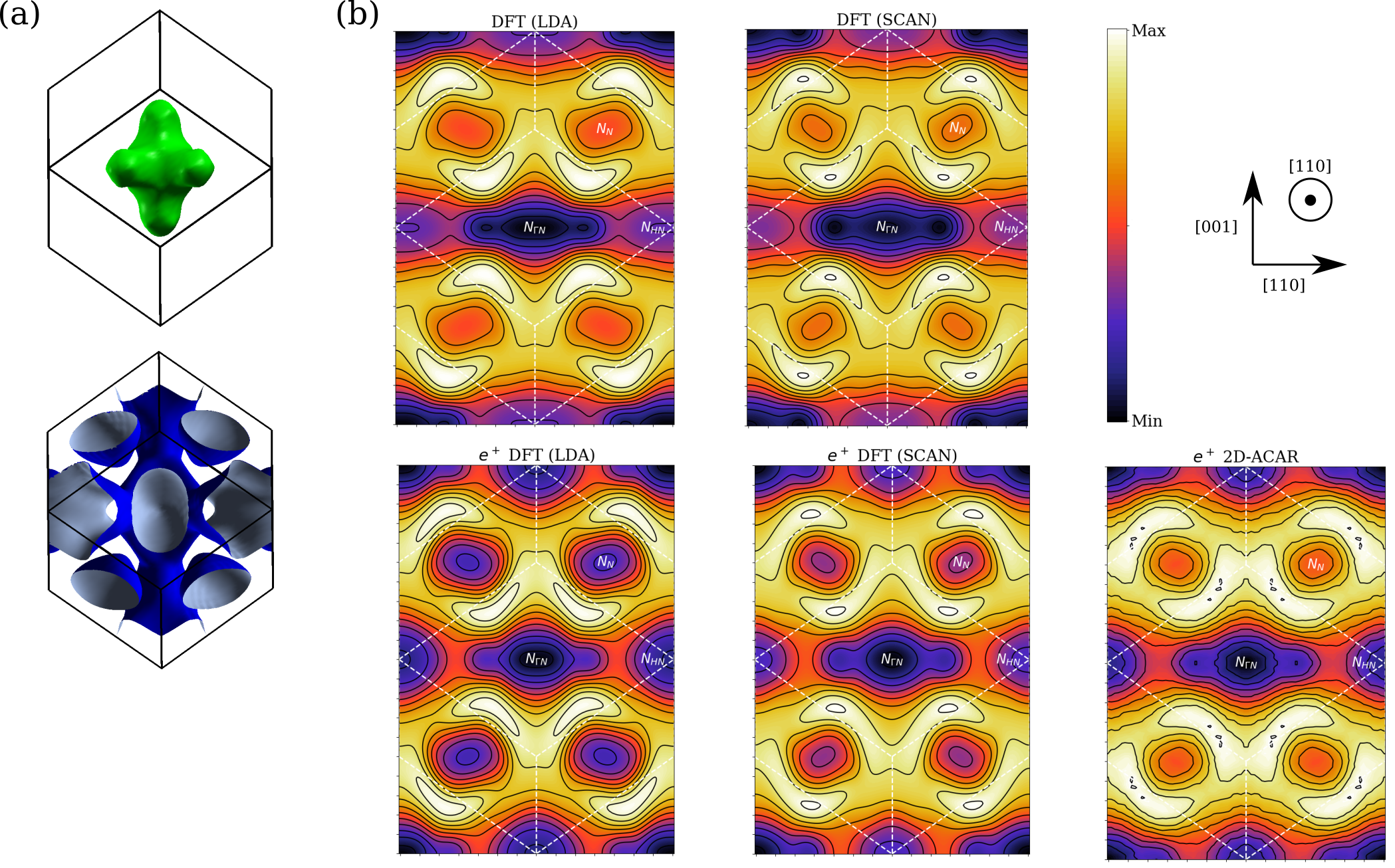}
	\caption{\label{fig:V_110_EMD} }
	a) The 3D Fermi surface of nonmagnetic V according to DFT-LDA calculations, rotated to the match the orientation of the projected 2D occupation densities.
	b) V 2D (once integrated) occupation densities projected along the [110] direction. Experimental results (bottom right) \cite{dugdale_bristol_2013,dugdale_application_1994} are compared to various theoretical calculations. The projection of the BCC Brillouin zone is indicated by the white dashed lines. The theoretical data has been convoluted with an appropriate experimental resolution function. The theoretical occupation densities on the bottom row include the influence of the positron on the measured electronic structure (using the enhancement of Drummond \textit{et al.} \cite{drummond_quantum_2011} for theoretical results), whereas plots on the top row do not.
\end{figure*}
The overall Fermi surface geometry of V is different to Cr and Mo, but the $N$ hole ellipsoid feature is shared.
A comparison to 2D-ACAR data for V \cite{dugdale_bristol_2013,dugdale_application_1994} is shown in Fig.~\ref{fig:V_110_EMD}.
V is a $3d$ (fourth period) transition metal, like Cr is, but the observable impact of the positron on the measured occupation density is smaller here than it is for Cr.
This could be because of the different Fermi surface geometry (and considerably larger N hole pockets), or it may be because the positron influences $p$ character states more strongly than $d$ character ones, and the band that creates the $N$ hole ellipsoids has a higher \% of $d$ character as it crosses the Fermi energy in V than in Cr.
In our calculations the $N$ hole pockets (at the $N_N$ point) of V are enlarged by the positron, and the apparent depth is also increased.
This means that the LDA result is actually worse than it appears to be when compared directly to the 2D-ACAR data; properly including the theoretical influence of the positron in the calculation increases the apparent difference between the LDA result and the experimental measurement.
The smaller $N$ hole pocket sizes in the SCAN calculations are more consistent with the experimental data than the ones for the LDA calculations, which are clearly too large.
The $N$ hole ellipsoids predicted by SCAN still seem a little bit larger than the experimental ones, which is consistent with the measured extremal areas being slightly too large (Fig.~\ref{fig:N_sizes_areas}).
Alternatively, the difference is small enough that it might just be a result of the approximate theoretical treatment of the positron.

\subsection{Other data (lattice constants and magnetic moments for Cr)}
\label{sec:other}

One of the major attractions of the SCAN MGGA is that it produces more accurate lattice constant predictions than LDA/GGA functionals when the main source of lattice constant error is the description of the region spanning the outermost core to innermost valence electrons, as it often is for metals \cite{sun_strongly_2015,fuchs_pseudopotential_1998,haas_insight_2009}.
The SCAN equilibrium lattice constant for Fe has previously been found to be closer to the experimental value than the PBE one, even though the magnetic moment is overestimated \cite{fu_applicability_2018}.
It has been noted that neither the LDA nor PBE GGA predicts the lattice constant of Cr accurately \cite{hafner_magnetic_2002} (assuming that the lattice constant of the nonmagnetic or commensurate antiferromagnetic state should match the experimental value), so we tested the SCAN MGGA here.
The predicted Cr lattice constant for different exchange-correlation functionals is shown in Table~\ref{tab:Cr_lat}. 
According to these calculations the SCAN MGGA does not seem to yield a more accurate lattice constant for Cr, with neither the nonmagnetic nor the commensurate antiferromagnetic lattice constants being closer to the experimental value than the PBE GGA predictions.
However, the true magnetic state at low temperature is a sinusoidally modulated incommensurate antiferromagnetic SDW state, which interpolates between the nonmagnetic solution at antinodes and the commensurate antiferromagnetic solution at nodes \cite{hafner_magnetic_2002}. 
It is therefore possible that the predicted lattice constant for the true magnetic state (an incommensurate antiferromagnetic SDW) would interpolate between the two listed values (2.83~-~2.95) and be closer to the experimental value.
\begin{table}
	\caption{\label{tab:Cr_lat} Table of predicted lattice constants for Cr, for nonmagnetic (NM) and commensurate antiferromagnetic (CAFM) calculations. The experimental lattice constant is a$_{expt}$~=~2.884~\r{A} \cite{wyckoff_crystal_1963}.}
	\begin{ruledtabular}
		\begin{tabular}{cccc}
			&LDA \cite{perdew_accurate_1992}&PBE \cite{perdew_generalized_1996}&SCAN \cite{sun_strongly_2015}
			\\
			\hline NM a$_{calc}$ (\r{A})&2.79&2.84&2.83
			\\
			CAFM a$_{calc}$ (\r{A})&2.79&2.86&2.95
			\\
		\end{tabular}
	\end{ruledtabular}
\end{table}

We note that the Fermi surface does show some level of dependence on the lattice constant, but we found that using the optimized lattice constant for a given functional, instead of the experimental one, does not particularly reduce the differences in Fermi surface between each functional, and the level of agreement with the experimental Fermi surface data is generally worsened (even when it is rescaled according to the relevant reciprocal lattice constant).

\begin{figure}
	\includegraphics[width=0.49\textwidth]{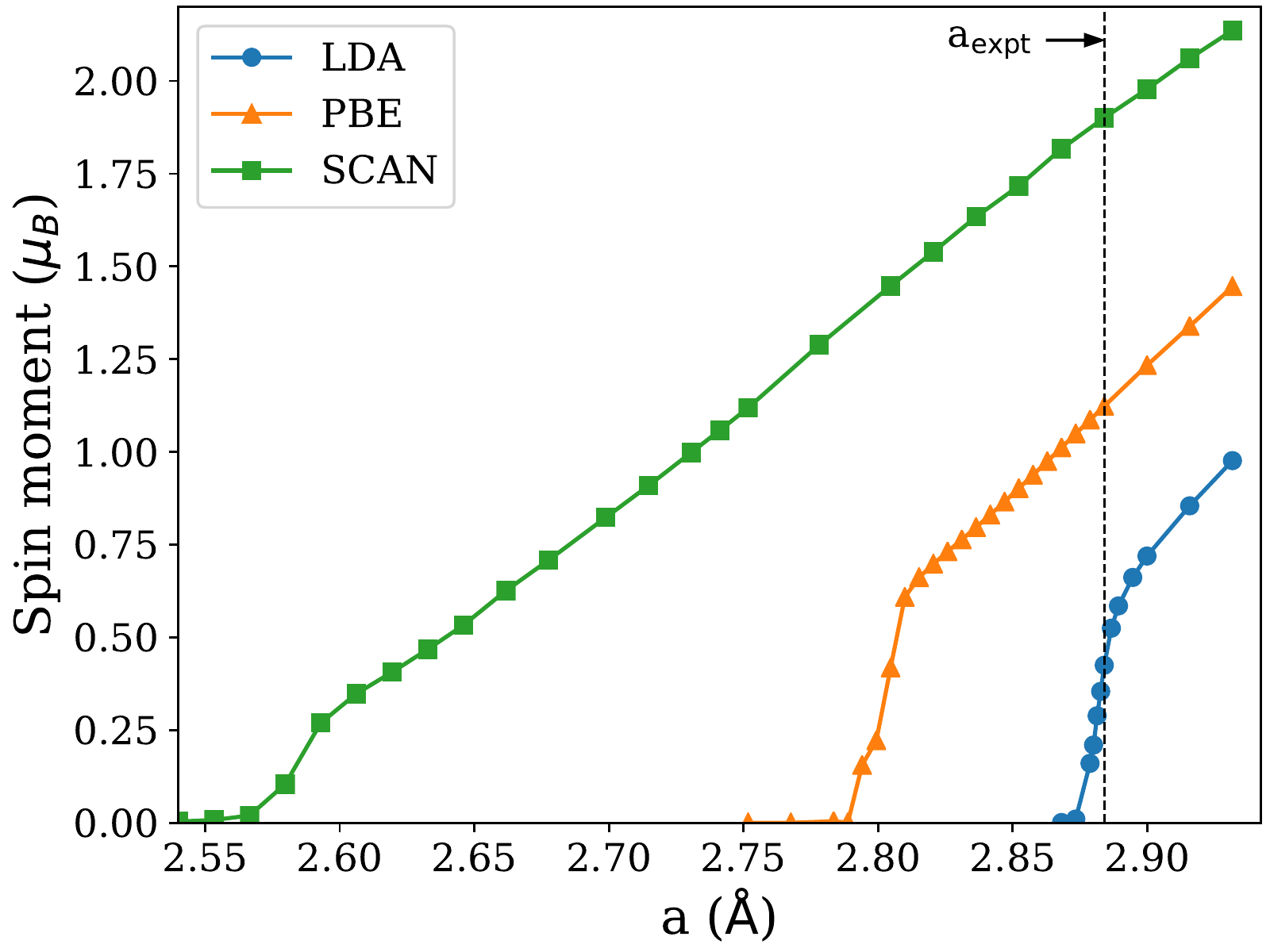}
	\caption{\label{fig:Cr_magmo} }
	Calculated Cr commensurate antiferromagnetic spin moments vs lattice constant for LDA, PBE GGA, and SCAN MGGA exchange-correlation density-functionals, according to Elk \cite{noauthor_elk_nodate}. The experimental lattice constant is indicated by the dashed line.
\end{figure}

\begin{figure}
	\includegraphics[width=0.49\textwidth]{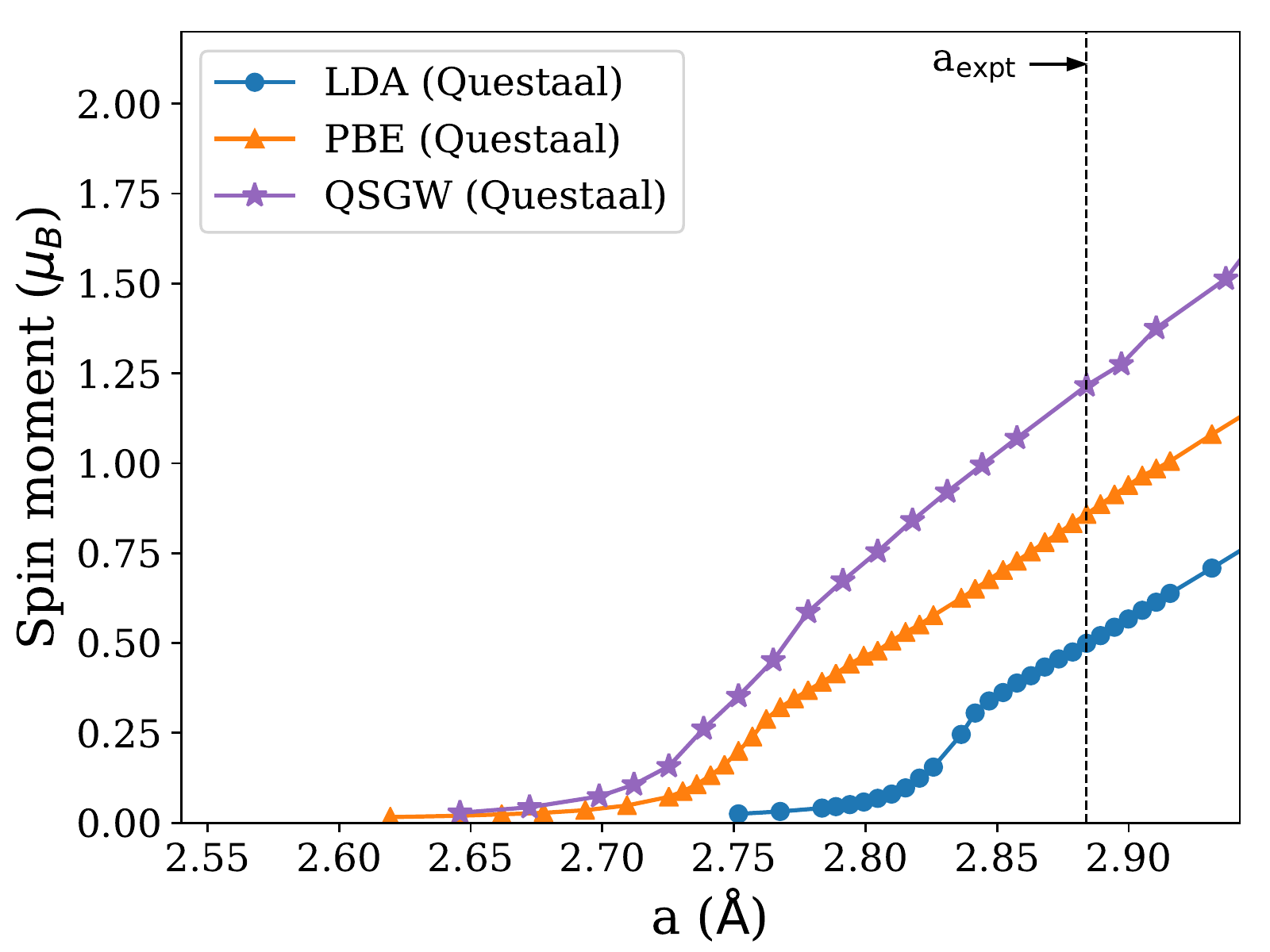}
	\caption{\label{fig:Cr_magmo_Q} }
	Calculated Cr commensurate antiferromagnetic spin moments vs lattice constant for LDA, PBE GGA, and QSGW calculations, all according to Questaal \cite{pashov_questaal_2020}. The experimental lattice constant is indicated by the dashed line.
\end{figure}

The predicted spin magnetic moments for calculations of the commensurate antiferromagnetic state of Cr (an SC Wigner-Seitz cell containing two Cr atoms) are listed in Table~\ref{tab:Cr_mag}.
The calculated relationship between magnetic moments and lattice constant for the LDA, PBE GGA, and SCAN MGGA exchange-correlation functionals is shown in Fig.~\ref{fig:Cr_magmo}.
For Questaal calculations, this relationship is shown in Fig.~\ref{fig:Cr_magmo_Q}.
Even for identical exchange-correlation approximations, there are clear differences between the two codes when it comes to this relationship.
The spin moments presented here for QSGW and SCAN should be gauged against the LDA and PBE results for the code that was used to calculate them.
This variation seems to be typical for Cr: there is an unusually wide variation in previously reported LDA and GGA results for the spin moments, as well as the dependence of the magnetism on the lattice constant \cite{hafner_magnetic_2002,cottenier_what_2002,soulairol_structure_2010,singh_magnetism_1992,marcus_mechanism_1998,kubler_spin-density_1980,hirai_magnetism_1997}.
In particular, we note that similar differences to those seen here were observed by Hafner \textit{et al.}, who also compared an LMTO code to LAPW and PAW ones \cite{hafner_magnetic_2002}.
The fact that different codes compute the moment in different ways may contribute.

It has previously been noted that the value of the theoretical commensurate antiferromagnetic moment matches the value for the theoretical amplitude of the antiferromagnetic SDW \cite{hafner_magnetic_2002}.
This allows us to compare our calculated antiferromagnetic moments to experimental measurements of the amplitude of the SDW.
The estimated amplitude of the SDW at T~=~4.2~K is $\mu_0~=~0.62~\mu_B$ per Cr atom, according to neutron-diffraction experiments \cite{arrott_neutron-diffraction_1967}.
This is about a factor of three smaller than the SCAN value that we obtained (1.9~$\mu_B$). 
The SCAN value is also close to double the size of the moment for the PBE GGA calculation at the same lattice constant.
(It is also worth noting that at the experimental lattice constant the difference in total energy between the non-spin-polarized and commensurate antiferromagnetic states is 230~meV/atom for SCAN, 24~meV/atom for PBE, and 1~meV/atom for LDA.)
The SCAN functional has previously been found to exhibit a stronger tendency toward magnetism than the LDA or GGA functionals, worsening agreement with various experimentally measured properties for Fe, Ni and Co \cite{isaacs_performance_2018,fu_applicability_2018,ekholm_assessing_2018,fu_density_2019}. 
According to these calculations Cr is no exception.
In fact, this is a larger overestimation, on both relative and absolute scales, than was obtained for SCAN calculations on Fe, Co, or Ni.
The QSGW estimate (1.22~$\mu_B$) is slightly larger than PBE, and therefore doesn't match the experimental value of 0.62~$\mu_B$ particularly well either.
But it has been argued that QSGW actually should systematically overpredict the magnetic moments of itinerant magnets \cite{sponza_self-energies_2017}.
\begin{table}
	\caption{\label{tab:Cr_mag} Table of predicted spin magnetic moments for the commensurate antiferromagnetic state of Cr. The experimental lattice constant is a$_{expt}$~=~2.884~\r{A} \cite{wyckoff_crystal_1963}.}
	\begin{ruledtabular}
		\begin{tabular}{ccccc}
			&LDA \cite{perdew_accurate_1992}&PBE  \cite{perdew_generalized_1996}&SCAN \cite{sun_strongly_2015}&QSGW\footnote{Questaal \cite{pashov_questaal_2020} calculation (others in this table are from Elk \cite{noauthor_elk_nodate}).}
			\\
			\hline
			$\mu$ ($\mu_B$) at a$_{expt}$&0.43&1.09&1.90&1.22
			\\
		\end{tabular}
	\end{ruledtabular}
\end{table}

\section{Discussion}

A pattern of discrepancies has been identified between LDA calculations and experimental measurements of the Fermi surfaces of the Group V and VI elemental transition metals.
It does not seem possible to attribute these discrepancies to limited experimental precision, or to other experimental challenges, because different experimental techniques highlight the same problems with the same Fermi surface feature (the $N$ hole ellipsoids), and because these same techniques are capable of achieving a higher level of agreement between experiment and theory in other metals.

Here we have provided evidence that the main factor in creating discrepancies is, not surprisingly, the exchange-correlation approximation.
Considering all of the experimental results for Cr and V, the $N$ hole ellipsoid sizes are consistently too large for LDA and PBE GGA calculations, but get smaller for SCAN and $GW$ calculations.
(It is also interesting to observe that, whereas the LDA and PBE GGA predictions are typically very similar to each other, the SCAN Fermi surface can be notably different and can be more similar to the QSGW result.)
For $G^{LDA}W^{LDA}$ calculations the sizes of these $N$ hole ellipsoids are much too small.
It is usually expected that $G^{LDA}W^{LDA}$ will be very accurate if the LDA starting point is a good approximation, but our results clearly show that the LDA starting point is not good enough for $GW$ calculations on Cr and V, and that introducing a level of self-consistency to the $GW$ calculation is important (the QSGW results are considerably more accurate).

The essential point is that for Cr and V the $N$ hole ellipsoid sizes depend strongly enough on the exchange-correlation approximation that it can explain the main part of experiment-theory discrepancies.
This also helps to explain why agreement between experiment and LDA predictions is better for the fifth period transition metal Mo; the LDA may actually be just as poor for Mo as it is for Cr, but the influence of the exchange-correlation approximation is smaller in Mo and as a result the approximation that is made to the exchange-correlation potential matters less.

The fact that different exchange-correlation approximations can produce a wide range of $N$ hole ellipsoid sizes may be related to the fact that this feature has a significant amount of $p$ character.
There are certain similarities between the way that the different exchange-correlation potentials describe this Fermi surface feature and the way that they predict the band gap sizes of $sp$-type semiconductors.
Consider Si for example, in which the bandgap is badly underpredicted by LDA and GGA calculations, slightly underpredicted by SCAN calculations \cite{yang_more_2016}, and slightly overpredicted by QSGW calculations \cite{vanschilfgaarde_quasiparticle_2006}.
The implication of this is that the Fermi surface discrepancies found in the Group V and VI transition metals may result from systematic problems with the LDA (and PBE GGA) when it comes to describing $s$ and $p$ character states.

In this work we have made the assumption that the nonmagnetic and paramagnetic states are similar enough that the differences between these cannot have created the observed discrepancies for V, Mo, and the paramagnetic state of Cr.
In V and Mo it is possible to obtain a very good level of agreement with the dHvA measurements (dHvA and 2D-ACAR) by choosing an appropriate exchange-correlation approximation, thus the assumption that the paramagnetic and nonmagnetic states have almost equivalent Fermi surfaces seems justified.
But Cr is more magnetically complex than its neighbours on the periodic table, exhibiting an antiferromagnetic SDW state, which is the only measurable state for one of the experimental techniques that has been analyzed here (dHvA).
Our calculations on the commensurate antiferromagnetic state of Cr demonstrate that this more complex magnetic state does have a different Fermi surface, but the $N$ hole ellipsoids, which are mainly of $p$ character, are not significantly affected (certainly not enough to explain the discrepancies between experiment and theory that are seen in Cr).
It is the electron ball sizes that depend on the magnetic state. 
Our calculations show that these reduce in volume when the $\Gamma$ electron and $H$ hole octahedra become magnetically gapped, compensating the larger hole volume of the octahedra at $H$ compared to the electron volume of the octahedra at $\Gamma$.
But differences in magnetic state cannot explain discrepancies between measurements and calculations of the sizes of the $N$ hole ellipsoids.
The dHvA measurements of the $N$ hole ellispoids of antiferromagnetic SDW Cr should therefore be representative of the paramagnetic state sizes.

We have shown that the difference between dHvA measurements and nonmagnetic LDA calculations on Cr is larger than previously recognized: both the electron balls and the $N$ hole ellipsoids are too large.
Part of the electron ball size discrepancy is corrected when the exchange-correlation approximation is changed so that the $N$ hole ellipsoid sizes are accurate.
But even when the $N$ hole ellipsoids are absent (in the $G^{LDA}W^{LDA}$ calculation) the predicted electron ball sizes are still slightly too large.
This seems to confirm that, although it is not as important a factor as the exchange-correlation approximation, proper treatment of the magnetic state is necessary to accurately predict the low temperature Fermi surface of Cr.

Remaining discrepancies between the calculations and 2D-ACAR measurements are also largest in Cr.
We also do not completely discount the possibility that, just as was the case for the dHvA comparison, there are remaining discrepancies related to the magnetic state. 
But this seems unlikely because the both the paramagnetic states of Mo and V seem adequately describable by nonmagnetic calculations, bearing out the expectation that the nonmagnetic and paramagnetic states should be equivalent to a good approximation.
Instead, any remaining discrepancies between the calculations and 2D-ACAR measurements mainly seem to be related to the imperfect theoretical treatment of the perturbative influence of the positron.
Our calculations show that the influence of the positron does seem to be particularly large in Cr.
It is clear from the [001] projection measurement that the SCAN calculation improves the Fermi surface description of Cr compared to LDA, and the SCAN prediction does also seem to improve on the LDA for the [110] projection, particularly for the $N$ hole ellipsoids (at the $N_N$ projection point), but differences do remain overall.
The fact that discrepancies are distinctly smaller for the [100] projection than [110] suggests that the k-dependent influence of the positron may still not have been fully accounted for in the theory (and this seems to be supported by previous maximum entropy filtering \cite{dugdale_fermiology_1998}).

To eliminate the positron influence, it is now possible to perform the Compton scattering measurements of the 2D [110] reconstruction to a higher resolution than the data shown here.
Higher resolution Compton scattering measurements would afford a better comparison of the $N$ hole ellipsoids, as well the Fermi surface in the rest of the zone.

For Mo and V, the positron influence on 2D-ACAR measurements appears to be smaller, but it is still clearly important to include it in the calculations so that the results of these can be better compared to the experimental results.
It is particularly important to include the theoretical influence of the positron for V; the difference between the LDA calculated $N$ hole ellipsoid sizes and the 2D-ACAR measurements is larger when the theoretical influence of the positron is included in the calculation, and the {\it apparent} accuracy of the LDA calculation that does not include the influence of the positron is therefore specious.

We have shown that Fermi surface discrepancies for the Group V and VI transition metals can largely be explained by the inadequacy of the exchange-correlation approximation, but it should be noted that even when the Fermi surface is well described, other measurable properties are not.
In particular, the magnetic properties of Cr are inadequately described, even by exchange-correlation approximations that improve the Fermi surface description.
SCAN dramatically overpredicts the magnetic moment, and the SDW that it forms for a 42 atom unit cell (that matches the period of the measured SDW) is not sinusoidally modulated (it is closer to a square wave).
In this respect SCAN is poorer than LDA and PBE GGA, which both do form this approximately sinusoidally modulated SDW.
Because of the clear and strong connection between lattice constant and magnetism in Cr, the lattice constant is also unlikely to be correctly predicted until the magnetic behaviour is adequately described.

It is unfortunate that the magnetic properties of Cr are not improved (and can actually be worsened) by the same exchange-correlation approximations that improve the Fermi surface description.
It may be that the Fermi surface discrepancies are mainly related to the description of $p$ character states, whereas the magnetic properties are mainly related to the $d$ character states.
It can be seen in Fig.~\ref{fig:CrMoV_bandstr_SCAN} that SCAN moves $d$ character bands further away from the Fermi energy (but only in the energy regions that are more than 1~eV from the Fermi energy - close to the Fermi energy the bands are not affected).
Photoemission data for V suggests that these $d$ bands are in reality closer to the Fermi energy than LDA predictions, not further from it \cite{peric_electronic_1995}.
Thus, there is experimental evidence that the $d$ band description by SCAN is worse than LDA.

A recent study of the ferromagnet Ni by Sponza \textit{et al.} used a combined QSGW+DMFT method (where DMFT is dynamical mean field theory \cite{georges_dynamical_1996}) to obtain very high level of agreement with various experimentally measurable properties simultaneously, including magnetic ones \cite{sponza_self-energies_2017}.
Sponza \textit{et al.} argue that first principles methods that do not include local spin fluctuation correlations, like QSGW, should overpredict the magnetic moment of itinerant magnets, including Ni and Cr.
They argue that all of the important correlations, including local spin fluctuations, are incorporated in QSGW+DMFT calculations and that this is why for Ni the magnetic moment, exchange-splitting, and $d$ band widths are all brought into good experimental agreement simultaneously by this method.
They found that QSGW does a reasonable job of predicting the Fermi surface of Ni, but other properties such as the magnetic moment and exchange-splitting are poorly predicted (the agreement with experiment is worse than it is for LDA calculations).
They argue that QSGW predicts these properties incorrectly, but for the right reason: the relevant correlations are completely omitted from the QSGW calculation.
A similar argument could be valid for Cr; it could be that local spin fluctuations are important, and that the predicted magnetic moment will therefore be too large unless these are taken into account in the calculation.
However, we cannot be confident that the hybridized $pd$ bands around the $N$ point, which are crucial to the accurate prediction of the Fermi surface of Cr, will be appropriately well treated with the double counting approximations used within DMFT, as these bands are sensitive to the exchange-correlation approximation.
The effect of the double counting approximation on the sizes of the Fermi surface sheets is too uncertain and this is why we did not include DMFT calculations here. 

In the future it could be interesting to check whether other exchange-correlation approximations, including hybrids, change the Fermi surface experiment-theory agreement for Cr and V as these SCAN and QSGW calculations do.
It would also be interesting to study the experiment-theory Fermi surface agreement for SCAN and QSGW for other metals - in cases where LDA/GGA agreement is unsatisfactory, as well as in cases where it is already satisfactory.
There are certainly plenty of other metals whose measured Fermi surfaces are already in very good agreement with LDA or GGA predictions, and ideally experiment-theory agreement ought not to be worsened in these cases. 
In cases where Fermi surface agreement between experiment and LDA/GGA calculations is not satisfactory, improvement may be possible if $p$ character states are important, as they are in the group V and VI transition metals tested here.
It is also desirable to find an exchange-correlation approximation that accurately predicts the Fermi surface as well as the other measurable properties (magnetism) of these metals simultaneously.

\section{Conclusion}

A pattern of discrepancies has been identified between LDA calculations and experimental measurements of the Group V and VI elemental transition metals.
These share a common Fermi surface feature, the N-centered hole ellipsoids, which are distinguished from the rest of the Fermi surface by the fact that they arise from a band that has a significant amount of $p$ character, rather than a $d$ character one.
The calculated size of these $p$-character-dominated $N$ hole ellipsoids is clearly dependent on the choice of exchange-correlation approximation for Cr and V, but less so for Mo.
This helps to explain the observation that agreement between experimental measurements and LDA calculations is better for Mo than it is for Cr.
We have also shown that the magnetic state of Cr does have an appreciable effect on its Fermi surface, although this is not as important a factor as the exchange-correlation approximation (the electron balls are affected by the magnetic state, but not the $p$ character $N$ hole ellipsoids, which are the biggest and most obvious source of discrepancy).

For the metals tested here, it is SCAN that most consistently predicts the Fermi surfaces accurately.
Further testing of SCAN and $GW$ (particularly QSGW) on other metals is warranted, especially the ones where $p$ character states play an important role.
Overall, this work serves as a reminder that it is not always necessary to look at metals with strongly correlated electronic behaviour to start to see problems with the LDA-predicted Fermi surface.

Although it most accurately predicts the Fermi surface, SCAN poorly predicts the magnetic properties of Cr (the predicted spin moment is 1.9 $\mu_B$, which is more than 3 times larger than the experimental value 0.62 $\mu_B$).
Ultimately, it would be more desirable to find an exchange-correlation approximation that accurately describes the Fermi surface along with the other measurable properties of these metals.

\section{Acknowledgement}

We are grateful to the developers of the Elk code, in particular J.~K.~Dewhurst and S.~Sharma. 
We are also grateful to the developers of the Questaal code, in particular to S.~Acharya and M.~van~Schilfgaarde for very valuable communications.
XCrySDen \cite{kokalj_xcrysdennew_1999} was used in the production of some figures.
The SKEAF code \cite{rourke_numerical_2012} was used to aid calculation of Fermi surface extremal areas.
The original research data are available at the University of Bristol data repository, \href{http://data.bris.ac.uk/data/}{data.bris}, at \href{https://doi.org/10.5523/bris.acsvraowhgga1ywx3qcboslo5}{http://doi.org/10.5523/bris.acsvraowhgga1ywx3qcboslo5} \cite{data_repository}.
Calculations were performed using the computational facilities of the Advanced
Computing Research Centre, University of Bristol
(\href{http://bris.ac.uk/acrc/}{http://bris.ac.uk/acrc/}).
E.~I.~Harris-Lee and A.~D.~N.~James acknowledge funding and support from the
UK Engineering and Physical Sciences Research Council (EPSRC). A.~D.~N.~James was supported by EPSRC Grant No.~EP/L015544/1 .

\end{document}